\documentclass[aps,prb,twocolumn,superscriptaddress,floatfix,longbibliography]{revtex4-2}

\usepackage{amsmath,amssymb} 
\usepackage{bm} 
\usepackage{graphicx} 
\usepackage{comment} 
\usepackage{textcomp} 
\usepackage{gensymb} 
\usepackage{booktabs}
\usepackage{makecell}
\usepackage{float}
\usepackage[left]{lineno}
\usepackage{lipsum, babel}

\usepackage{dcolumn} 
\usepackage{bm}     
\usepackage{amsfonts}
\usepackage[bookmarks=false,colorlinks,citecolor=blue]{hyperref}
\hypersetup{colorlinks=true,linkcolor=blue,filecolor=blue,citecolor = blue, urlcolor=blue}\usepackage{amsmath}
\newcommand{\angstrom}{\textup{\AA}}
\usepackage[version=4]{mhchem}
\usepackage{siunitx}

\usepackage{mathtools}

\DeclarePairedDelimiterX\braket[2]{\langle}{\rangle}{#1\,\delimsize\vert\,\mathopen{}#2}

\usepackage{enumitem}
\setlist{noitemsep,leftmargin=*,topsep=0pt,parsep=0pt}

\usepackage{xcolor} 
\definecolor{lightgray}{gray}{0.6}
\definecolor{medgray}{gray}{0.4}

\newif\ifptitle
\newif\ifpnumber
\newcounter{para}

\ptitletrue  
\pnumbertrue  

\newcommand{\mytitle}{Topotactic oxidation of Ruddlesden–Popper nickelates reveals new structural family: oxygen-intercalated layered perovskites}

\begin{document}

\title{\mytitle}
\author{Dan Ferenc Segedin}
\affiliation{Department of Physics, Harvard University, Cambridge, MA 02138, USA}
\author{Jinkwon Kim}
\affiliation{Department of Materials Science and Engineering, Cornell University, Ithaca, New York 14853, USA}
\author{Harrison LaBollita}
\affiliation{Department of Physics, Arizona State University, Tempe, AZ 85287, USA}
\affiliation{Center for Computational Quantum Physics, Flatiron Institute, 162 5th Avenue, New York, New York 10010, USA.}
\author{Nicole K. Taylor}
\affiliation{School of Engineering and Applied Science, Harvard University, Cambridge, MA 02138, USA}
\author{Kyeong-Yoon Baek}
\affiliation{Department of Physics, Harvard University, Cambridge, MA 02138, USA}
\author{Suk Hyun Sung}
\affiliation{The Rowland Institute at Harvard, Harvard University, Cambridge, MA 02138, USA}
\author{Ari B. Turkiewicz}
\affiliation{Department of Physics, Harvard University, Cambridge, MA 02138, USA}
\author{Grace A. Pan}
\affiliation{Department of Physics, Harvard University, Cambridge, MA 02138, USA}
\author{Abigail Y. Jiang}
\affiliation{School of Engineering and Applied Science, Harvard University, Cambridge, MA 02138, USA}
\author{Maria Bambrick-Santoyo}
\affiliation{Department of Physics, Harvard University, Cambridge, MA 02138, USA}
\affiliation{Department of Electrical Engineering and Computer Science, Massachusetts Institute of Technology, Cambridge, MA 02138, USA}
\author{Tobias Schwaigert}
\affiliation{Department of Materials Science and Engineering, Cornell University, Ithaca, New York 14853, USA}
\author{Casey K. Kim}
\affiliation{Department of Materials Science and Engineering, Cornell University, Ithaca, New York 14853, USA}
\author{Anirudh Tenneti}
\affiliation{Smith School of Chemical and Biomolecular Engineering, Cornell University, Ithaca, New York 14853, USA}
\author{Alexander J. Grutter}
\affiliation{%
 NIST Center for Neutron Research, National Institute of Standards and Technology, Gaithersburg, MD 20899, USA
}%
\author{Shin Muramoto}
\affiliation{%
Material Measurement Laboratory, National Institute of Standards and Technology, Gaithersburg, MD 20899, USA
}%
\author{Alpha T. N'Diaye}
\affiliation{Advanced Light Source, Lawrence Berkeley National Laboratory, Berkeley, CA 94720, USA}
\author{Ismail El Baggari}
\affiliation{The Rowland Institute at Harvard, Harvard University, Cambridge, MA 02138, USA}
\author{Donald A. Walko}
\affiliation{Advanced Photon Source, Argonne National Laboratory, Lemont, IL 60439, USA}
\author{Hua Zhou}
\affiliation{Advanced Photon Source, Argonne National Laboratory, Lemont, IL 60439, USA}
\author{Charles M. Brooks}
\affiliation{Department of Physics, Harvard University, Cambridge, MA 02138, USA}
\author{Antia S. Botana}
\affiliation{Department of Physics, Arizona State University, Tempe, AZ 85287, USA}
\author{Darrell G. Schlom}
\affiliation{Department of Materials Science and Engineering, Cornell University, Ithaca, New York 14853, USA}
\affiliation{Kavli Institute at Cornell for Nanoscale Science, Ithaca, New York 14853, USA}
\affiliation{Leibniz-Institut für Kristallzüchtung, Max-Born-Str. 2, 12489 Berlin, Germany}
\author{Julia A. Mundy}
\thanks{\href{mailto:mundy@fas.harvard.edu}{mundy@fas.harvard.edu}}
\affiliation{Department of Physics, Harvard University, Cambridge, MA 02138, USA}
\affiliation{School of Engineering and Applied Science, Harvard University, Cambridge, MA 02138, USA}
\date{}

\begin{abstract}
Layered perovskites such as the Dion-Jacobson, Ruddlesden-Popper, and Aurivillius families host a wide range of correlated electron phenomena, from high-temperature superconductivity to multiferroicity. Here we report a new family of layered perovskites, realized through topotactic oxygen intercalation of La$_{n+1}$Ni$_{n}$O$_{3n+1}$ ($n=1-4$) Ruddlesden–Popper nickelate thin films grown by ozone-assisted molecular-beam epitaxy. Post-growth ozone annealing induces a large $c$-axis expansion – 17.8\% for La$_{2}$NiO$_{4}$ ($n=1$) – that monotonically decreases with increasing $n$. Surface X-ray diffraction coupled with Coherent Bragg Rod Analysis reveals that this structural expansion arises from the intercalation of approximately 0.7 oxygen atoms per formula unit into interstitial sites within the rock salt spacer layers. The resulting structures exhibit a spacer layer composition intermediate between that of the Ruddlesden-Popper and Aurivillius phases, defining a new class of layered perovskites. Oxygen-intercalated nickelates exhibit metallicity and significantly enhanced nickel-oxygen hybridization, a feature linked to high-temperature superconductivity. Our work establishes topotactic oxidation as a powerful synthetic approach to accessing highly oxidized, metastable phases across a broad range of layered oxide systems, offering new platforms to tune properties via spacer-layer chemistry.

\end{abstract}

\maketitle

Perovskite oxides ($AB$O$_3$, $A$, $B$ = cation) exhibit an extraordinary range of emergent phenomena including superconductivity, multiferroicity, and colossal magnetoresistance \cite{Sleight1975_BaBiO3_SC,Wang2003_BFO, Jin1994_CMR}. These materials also adopt naturally layered structures, such as the Dion–Jacobson \cite{Dion1981, jacobson1985interlayer}, Ruddlesden-Popper \cite{Balz1955K2NiF4,RuddlesdenPopper1957_K2NiO4typecompounds}, and Aurivillius \cite{aurivillius1949mixed, aurivillius2, aurivillius3, aurivillius4} families. Layered perovskites can be represented by the formula ($A_{x}$O$_{y}$)($A_{n-1}$$B_{n}$O$_{3n+1}$), comprising $n$ perovskite-like layers, ($A_{n-1}$$B_{n}$O$_{3n+1}$), separated by family-specific spacer layers: $A$ in Dion-Jacobson, $A_{2}$ in Ruddlesden-Popper, and $A_{2}$O$_{2}$ in Aurivillius compounds. These spacer layers modulate the dimensionality of the structure, from quasi-two-dimensional ($n=1$) to three-dimensional in the perovskite limit ($n = \infty$). Recently, Ruddlesden-Popper nickelates have attracted significant attention following the discovery of high-temperature superconductivity under hydrostatic pressure in the bilayer and trilayer members, La$_3$Ni$_2$O$_7$ and La$_4$Ni$_3$O$_{10}$ \cite{Sun2023_SCLa327, Zhu2024La4310SC}. Epitaxial strain was subsequently shown to stabilize superconductivity above 40 K in La$_{3}$Ni$_{2}$O$_{7}$ thin films, providing a pathway to realize high-temperature superconductivity in Ruddlesden-Popper nickelates at ambient pressure \cite{Ko2024Nature, Zhou2024La327SC}.

\begin{figure*}
    \centering
    \includegraphics[width = 2\columnwidth]{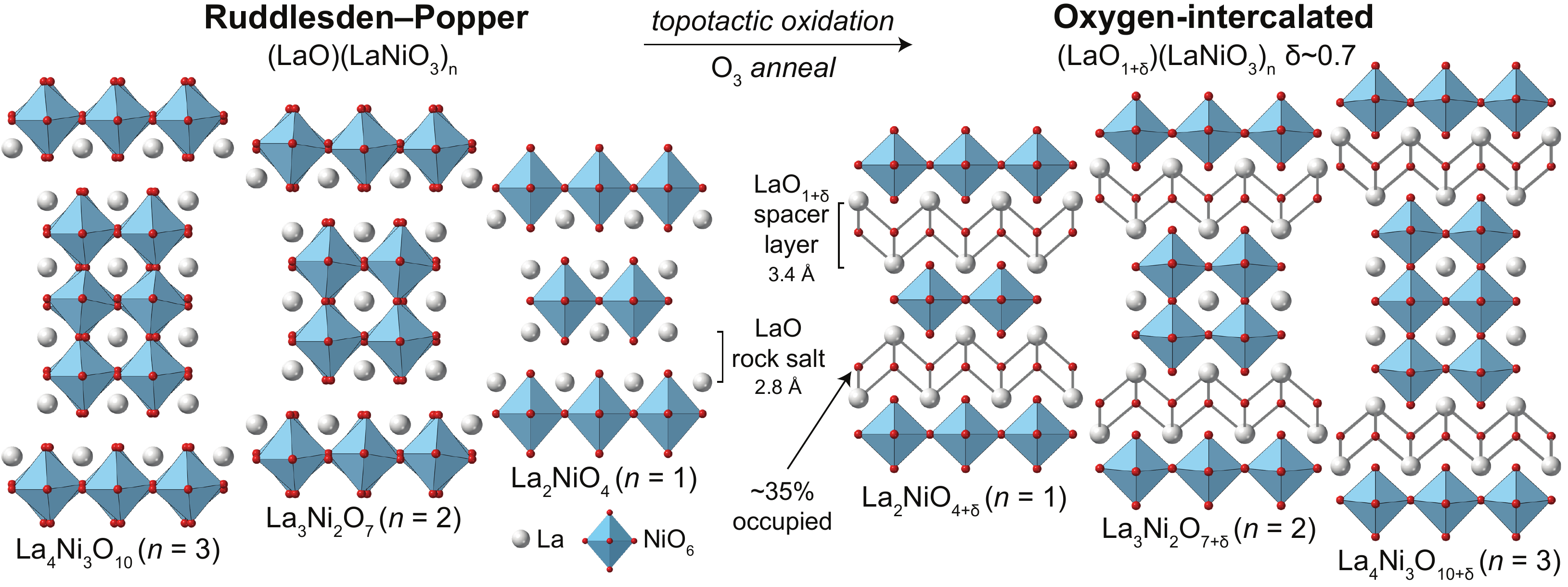}
    \caption{\textbf{Topotactic oxidation of Ruddlesden-Popper nickelates.} Schematic crystal structures of Ruddlesden–Popper \textbf{(left)} and oxygen-intercalated layered \textbf{(right)} nickelates. Topotactic oxidation introduces approximately 0.7 oxygen atoms per formula unit into interstitial sites within the rock salt layer, driving an expansion of the spacer layer. Approximately 35\% of the interstitial sites are occupied by oxygen intercalants. 
    }
    \label{fig1_structures}
\end{figure*}

In addition to dimensionality and epitaxial strain, soft-chemical synthetic methods have demonstrated remarkable tunability of oxide properties through controlled modification of the oxygen content \cite{Hayward2013}. For example, topotactic reduction of Ruddlesden-Popper nickelates via oxygen deintercalation yields a distinct structural family: the square-planar nickelates, $A_{n+1}$Ni$_{n}$O$_{2n+2}$ or ($A$NiO$_{2}$)$_{n}$($A$O$_{2}$)$^{-}$. These compounds feature $n$ $A$NiO$_{2}$ square-planar layers separated by $A$O$_{2}$ fluorite-like layers, and host nickel in a reduced $\sim$$1+$ oxidation state. Isoelectronic ($\sim$$d^{9}$) and isostructural (square-planar) with the cuprates, reduced nickelates feature superconductivity with notable similarities and differences to the high-$T_{c}$ copper-oxides \cite{LiNat2019, PanNatMat2021}. In principle, Ruddlesden–Popper nickelates also support topotactic oxidation via oxygen intercalation: nickel can access oxidation states up to 4+ and rock salt spacer layers contain interstitial sites capable of hosting excess oxygen \cite{Yilmaz2024NPJ_Ni4+,AguaderoJourMatChem2006_La2NiO4p34}. Yet, a topotactic structural transformation driven by oxygen intercalation has not been realized in any Ruddlesden-Popper oxide to date. 

Here, we uncover a new family of layered perovskite materials through topotactic oxygen intercalation of Ruddlesden-Popper nickelate thin films. As shown in Fig.\ \ref{fig1_structures}, oxygen-intercalated nickelates incorporate approximately 0.7 oxygen atoms per formula unit in the spacer layer, inducing a dramatic expansion in the out-of-plane lattice constant relative to their Ruddlesden-Popper precursors. This new crystal structure is an intermediary between the Ruddlesden-Popper and Aurivillius phases. We demonstrate that oxygen intercalation alters the crystal structure of the perovskite layers, the nickel $d$-electron filling, nickel-oxygen hybridization, and electrical transport properties. Oxygen-intercalated nickelates establish a versatile new platform to tune nickelate properties via spacer layer chemistry. More broadly, our results highlight ozone annealing as a powerful soft-chemical approach to achieve topotactic oxygen intercalation in layered oxides.

We note that ozone annealing is essential to fully oxidize and stabilize superconductivity in bi-layer (La,Pr)$_{3}$Ni$_{2}$O$_{7}$ films; however, excessive oxidation can destabilize the Ruddlesden-Popper phase. Prior studies have reported that strongly oxidizing conditions can induce the formation of a `perovskite' phase in (La,Pr)$_{3}$Ni$_{2}$O$_{7}$ films \cite{Ko2024Nature, Liu2025ozone}. We propose that this destabilized `perovskite' phase may be related to the oxygen-intercalated compounds reported here.

\begin{figure*}
    \centering
    \includegraphics[width = 2\columnwidth]{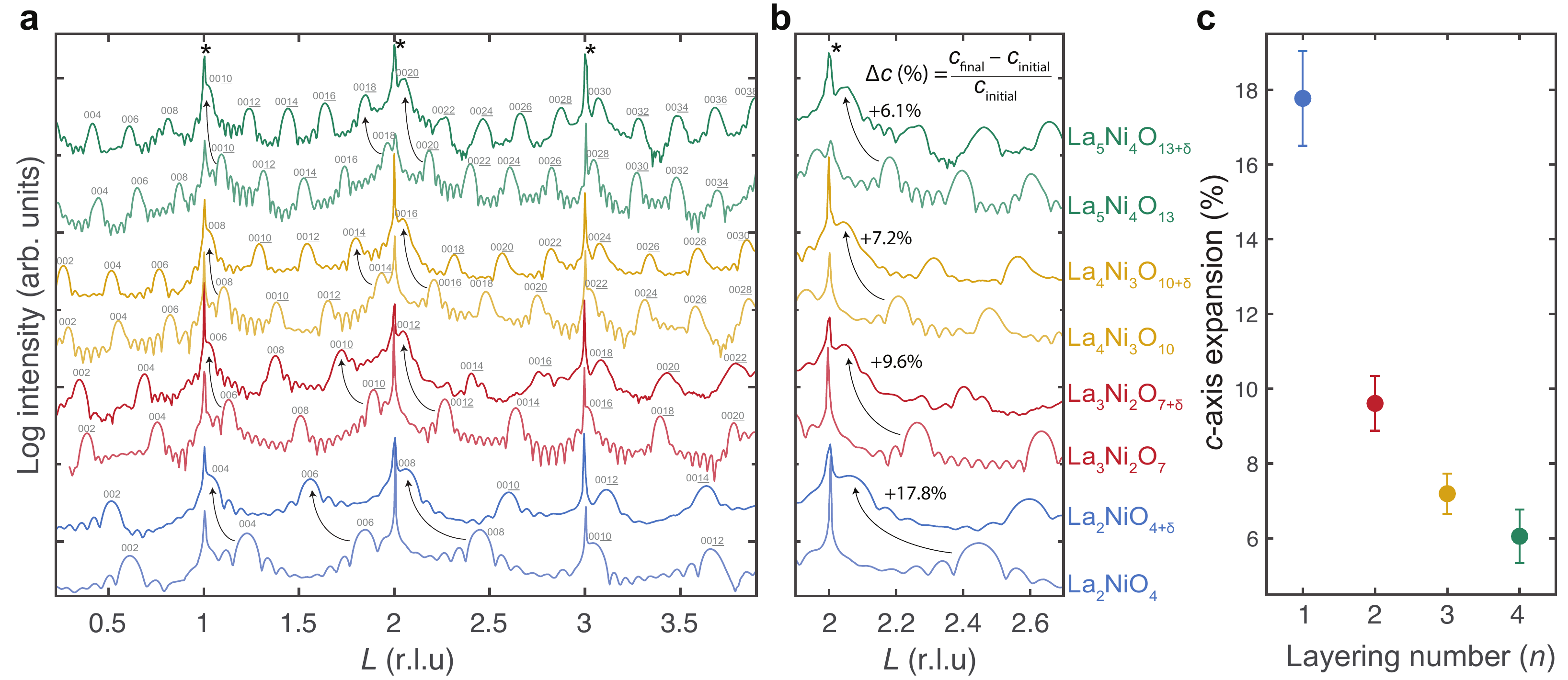}
    \caption{\textbf{Structural characterization via X-ray diffraction of La$_{n+1}$Ni$_{n}$O$_{3n+1}$ ($n=1-4$) films before and after ozone treatment.} \textbf{a,} Synchrotron X-ray diffraction scans along the (0,0,$L$) specular crystal truncation rod. Scans are vertically offset for clarity. Lab-based X-ray scans are available in Fig.\ \ref{lab_xray}. Reciprocal space units are normalized to NdGaO$_{3}$ (110) substrate Bragg peaks, marked with asterisks. Arrows indicate peak shifts following ozone treatment. \textbf{b,} Zoom-in view of scans in \textbf{(a)}. \textbf{c,} Percentage increase in $c$-axis lattice constant after oxidation. Lattice constants are provided in Table \ref{tab:caxis}.
    }
    \label{fig2_00L_scans}
\end{figure*}

\begin{figure*}
    \centering
    \includegraphics[width = 2\columnwidth]{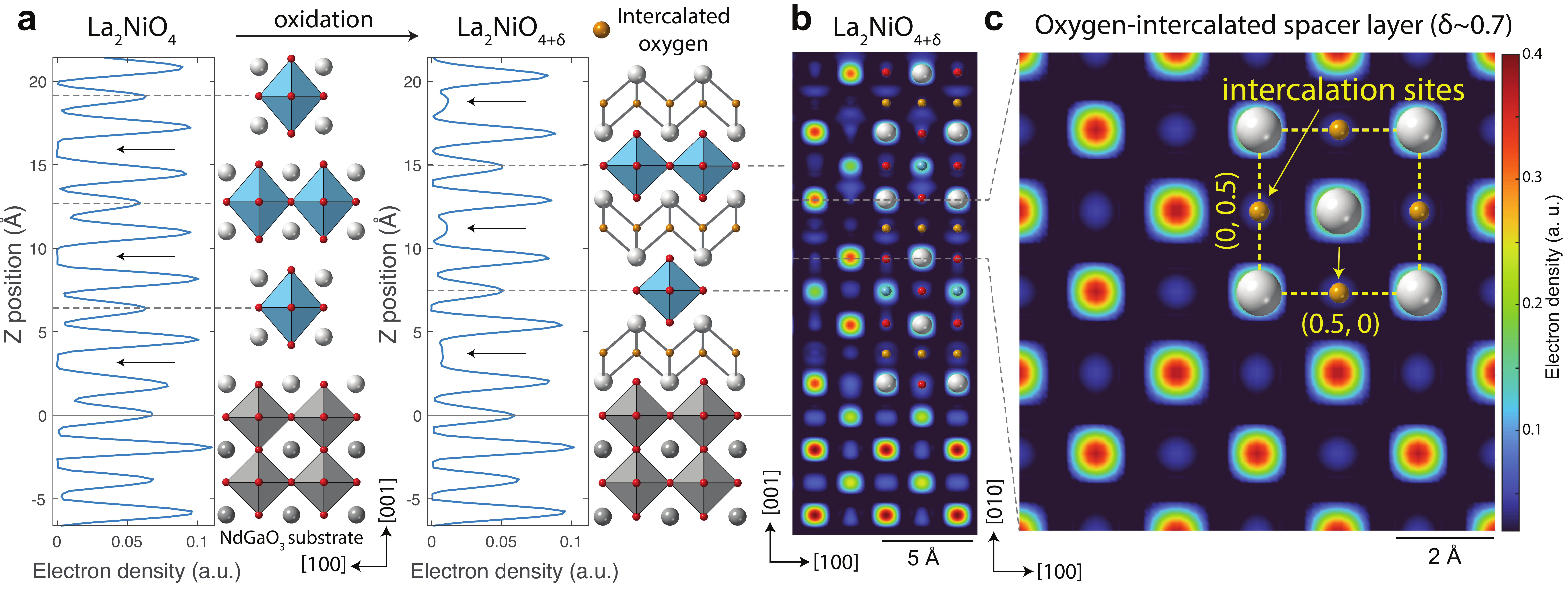}
    \caption{\textbf{Structural characterization via Coherent Bragg Rod Analysis (COBRA).} \textbf{a,} One-dimensional planar-averaged electron density profiles of as-grown \textbf{(left)} and oxidized \textbf{(right)} La$_{2}$NiO$_{4}$ obtained by COBRA (Fig.\ \ref{cobra_00L_fits}). Arrows indicate the spacer layers, which expand and show increased electron density after oxidation. Solid and dashed lines denote the substrate-film interface and NiO$_{2}$ layers, respectively. \textbf{b,} Projected two-dimensional electron density along the in-plane [010] direction, generated by combining cuts at ($a$,$b$) = (0,0) and (0,0.5). \textbf{c,} Projected two-dimensional in-plane electron density of the spacer layer, constructed from cuts through center of the spacer layer and two adjacent LaO planes, as indicated by dashed lines in \textbf{(b)}. The pseudocubic unit cell is represented by the yellow dashed square.}
    \label{fig3_cobra}
\end{figure*}

\section{Results and discussion}

\subsection{Structural characterization}

We synthesize Ruddlesden-Popper La$_{n+1}$Ni$_n$O$_{3n+1}$ ($n=1-4$) thin films on NdGaO$_{3}$ (110) substrates using ozone-assisted molecular-beam epitaxy (MBE). Topotactic oxidation is subsequently carried out via ozone annealing (see Methods). We employ synchrotron surface X-ray diffraction to characterize the film lattice and atomic structure before and after oxidation. Figure \ref{fig2_00L_scans}a,b, presents (0,0,$L$) specular crystal truncation rods (CTRs), which are sensitive to out-of-plane structural order (lab-based X-ray diffraction data provided in Fig.\ \ref{lab_xray}). Upon oxidation, all films exhibit pronounced leftward peak shifts, indicative of an increase in the out-of-plane ($c$-axis) lattice constant, as illustrated in Fig.\ \ref{fig1_structures}. The single-layer La$_{2}$NiO$_{4}$ film, for example, exhibits a 17.8\% $c$-axis expansion from 12.68 $\angstrom$ to 14.93 $\angstrom$ (see Table \ref{tab:caxis} for lattice constants). The fractional expansion, $\Delta c / c$, monotonically decreases with increasing $n$, indicating that expansion occurs mostly within the spacer layers (Fig.\ \ref{fig2_00L_scans}c). The absolute expansion, $\Delta c$, is approximately 2.1 $\angstrom$ for all values of $n$. 

Reciprocal space mapping indicates that the films remain epitaxially strained to the substrate after the topotactic transformation (Fig.\ \ref{214_rsm}). Topotactic oxidation proceeds similarly on other substrates, including LaAlO$_{3}$ (100) and SrTiO$_{3}$ (Fig.\ \ref{La327_strain}), and with different $A$-site cations such as neodymium (Fig.\ \ref{Nd214_xrd}). In all cases, the oxidized films exhibit topotactic expansion and remain strained, demonstrating that oxygen-intercalated nickelates are strain- and $A$-site-tunable. Furthermore, the oxidation process is reversible: annealing the oxidized film in air restores the parent Ruddlesden–Popper phase (Fig.\ \ref{air_anneal}).

\begin{figure*}
    \centering
    \includegraphics[width = 1.7\columnwidth]{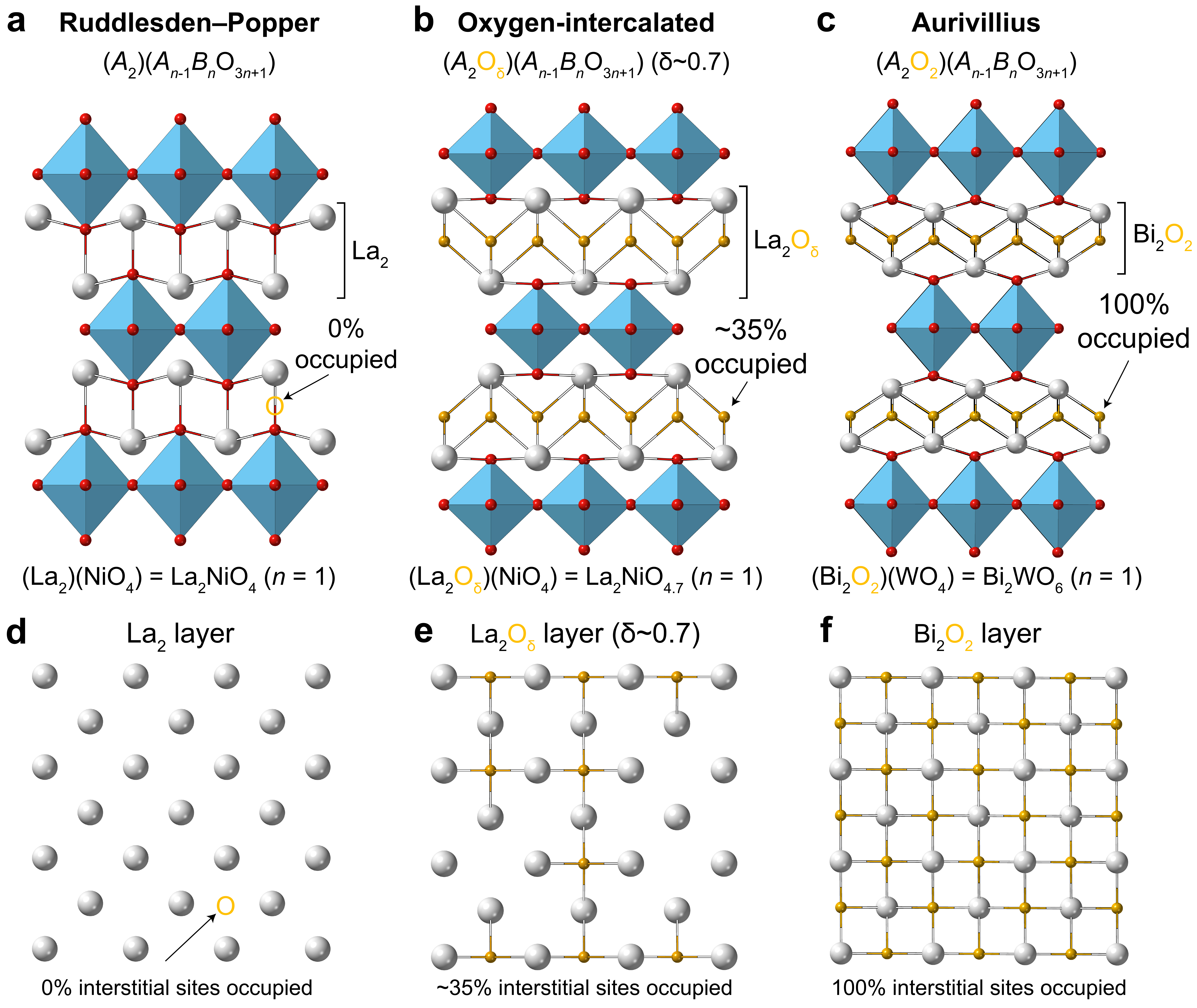}
    \caption{\textbf{Families of layered perovskites.} \textbf{a,b,c,} Schematic crystal structures of Ruddlesden–Popper \textbf{(a)}, oxygen-intercalated \textbf{(b)}, and Aurivillius \textbf{(c)} phases. \textbf{d,e,f,} Spacer layer structures for La$_{2}$ \textbf{(d)}, La$_{2}$O$_{\delta}$ \textbf{(e)}, and Bi$_{2}$O$_{2}$ \textbf{(f)}. Interstitial sites are marked with a gold circle in (a) and (d). Interstitial and apical oxygen atoms are gold and red, respectively.}
    \label{fig5_rp_vs_aurivillius}
\end{figure*}

To investigate the structural origin of the large $c$-axis expansion, we employ coherent Bragg rod analysis (COBRA) to the measured CTRs, including both specular and non-specular Bragg rods \cite{Yacoby2002COBRA, Zhou2010}. COBRA analysis of (0,0,$L$) specular CTRs (Fig.\ \ref{cobra_00L_fits}) yields an in-plane averaged electron density profile of the film. Figure \ref{fig3_cobra}(a) presents the reconstructed electron density for the parent and oxidized La$_{2}$NiO$_{4}$ film. In the parent compound, the rock salt spacer layers exhibit a characteristic double-peak structure separated by a gap. Upon oxidation, the spacer layer expands from 2.77 $\angstrom$ to 3.41 $\angstrom$  (+23.1\%). The perovskite layers also expand but by a smaller amount: 3.53 $\angstrom$ to 3.95 $\angstrom$ (+11.9\%). Furthermore, additional electron density emerges within the expanded spacer layer, indicating oxidation-induced intercalation. These results demonstrate that ozone annealing La$_{2}$NiO$_{4}$ induces oxygen intercalation of the rock salt spacer layer, driving an expansion in the $c$-axis lattice constant. Density functional theory (DFT) calculations of oxygen-intercalated La$_{n+1}$Ni$_{n}$O$_{3n+1+\delta}$ ($n$ = 1 - 3, $\delta=1$) qualitatively reproduce the experimentally observed structural expansion (Supplementary Note 2, Table \ref{tab:dftgeometry}). 

\begin{figure*}
    \centering
    \includegraphics[width = 2\columnwidth]{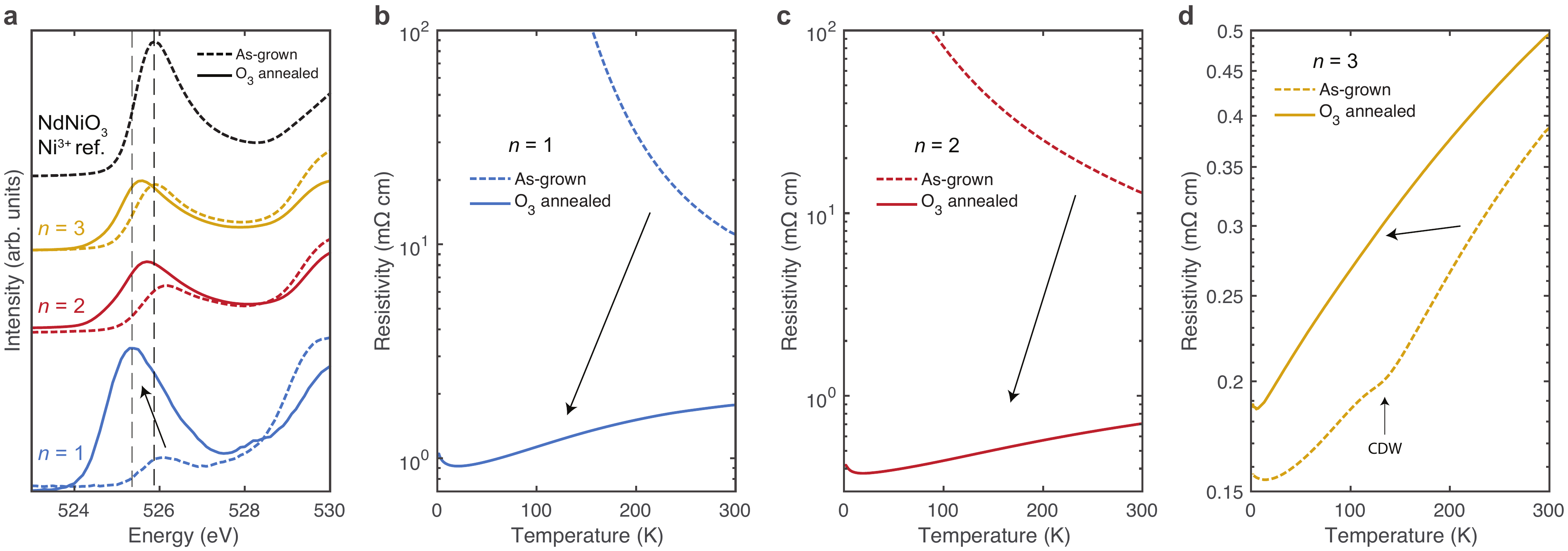}
    \caption{\textbf{Electronic structure and transport characterization of La$_{n+1}$Ni$_{n}$O$_{3n+1}$ films before and after oxidation.} \textbf{a,} X-ray absorption spectra at oxygen-$K$ edge. The vertical dashed lines indicate the pre-peak positions of the nickel 3+ reference and the oxidized La$_{2}$NiO$_{4+\delta}$ film. \textbf{b,c,d,} Temperature-dependent resistivity before and after ozone treatment for La$_{2}$NiO$_{4}$ ($n$ = 1) \textbf{(b)}, La$_{3}$Ni$_{2}$O$_{7}$ ($n$ = 2) \textbf{(c)}, and La$_{4}$Ni$_{3}$O$_{10}$ ($n$ = 3) \textbf{(d)}.}
    \label{fig4_xas_transport}
\end{figure*}

We characterize the composition and structure of the intercalated spacer layer using 3-D COBRA, which reconstructs the three-dimensional electron density from both specular and nonspecular CTRs (Fig.\ \ref{La215_nonspecular_CTRs}). A projected two-dimensional electron density map of oxidized La$_{2}$NiO$_{4+\delta}$ is presented in Figure \ref{fig3_cobra}(b). Pockets of electron density are visible within the expanded spacer layer, corresponding to intercalated oxygen atoms. The in-plane projection of the spacer layer in Fig.\ \ref{fig3_cobra}(c) reveals that the intercalants occupy interstitial sites ($a$,$b$) = (0, 0.5) and (0.5, 0). The projection along the [110] pseudocubic direction shows additional intercalation sites at ($a$,$b$) = (0, 0) (Fig.\ \ref{cobra_110_spacer}). Integration of the electron density at the interstitial sites yields approximately 5.8 electrons, corresponding to $\delta\sim$ 0.7 $\pm$ 0.1 interstitial oxygen atoms.

Half-order CTRs, which are sensitive to doubling periodicity arising from oxygen octahedral rotations \cite{May2010PRB}, reveal that oxidation nearly eliminates octahedral rotations in the $n=2$ compound and partially suppresses them in the $n=3$ film (Fig.\ \ref{n2_n3_halforder}). These observations suggest that topotactic oxidation offers a means to modulate oxygen octahedral structural frameworks, a structural degree of freedom linked to high-temperature superconductivity in bi-layer and tri-layer Ruddlesden-Popper nickelates \cite{Sun2023_SCLa327, Zhu2024La4310SC}.

Oxygen-intercalated layered nickelates appear to form a new family of layered perovskites, structurally intermediate between the Ruddlesden–Popper and Aurivillius phases. As illustrated in Fig.\ \ref{fig5_rp_vs_aurivillius}, these canonical phases are distinguished by the composition of the spacer layers that separate the octahedra: $A_{2}$ in Ruddlesden–Popper compounds and $A_{2}$O$_{2}$ in Aurivillius phases. The $A_{2}$ layer forms a square lattice of interstitial sites, which are empty in the Ruddlesden–Popper structure (Fig.\ \ref{fig5_rp_vs_aurivillius}d) and fully occupied in Aurivillius compounds (Fig.\ \ref{fig5_rp_vs_aurivillius}f). In contrast, the oxygen-intercalated nickelates incorporate approximately 0.7 interstitial oxygen atoms per formula unit (Fig.\ \ref{fig5_rp_vs_aurivillius}b), yielding a spacer layer of nominal composition $A_{2}$O$_{0.7}$ and 35\% interstitial site occupancy (Fig.\ \ref{fig5_rp_vs_aurivillius}e). 

To our knowledge, this work represents the first demonstration of a topotactic structural transformation in any Ruddlesden-Popper oxide induced by oxygen intercalation. Although oxygen intercalation has been observed in various Ruddlesden-Popper systems, the levels of intercalation reported are significantly lower than the $\delta\sim0.7$ achieved here — for example, $\delta=0.34$ in La$_{2}$NiO$_{4+\delta}$ \cite{AguaderoJourMatChem2006_La2NiO4p34}, $\delta=0.32$ in La$_{2}$CoO$_{4+\delta}$ \cite{AguaderoNaturforshung_La2CoO4.32}, and $\delta=0.18$ La$_{2}$CuO$_{4+\delta}$ \cite{JorgensenPRB1988}. Furthermore, these oxygen-rich systems do not exhibit structural expansion relative to the stoichiometric ($\delta=0$) phase. Our results point to a threshold oxygen intercalation — between $\delta\sim$ 0.3 and 0.7 — required to drive rock salt layer expansion. 

Topotactic structural transformations have previously been observed in other Ruddlesden-Popper oxides following fluorination \cite{Slater2001fluorine, Nowroozi2017LaSrMnO4F2, Wissel2020ChemMat_Sr3Ti2O5F4}, nitrogenation \cite{hong2025_Sr2CoO4_oxidation}, and hydration \cite{Slater2001Ba2ZrO3F2xH2O, Pelloquin2005, Berry2008Ba2SnO4}. Notably, the fluorinated nickelate La$_{2}$NiO$_{3}$F$_{2}$ shows no $c$-axis expansion relative to the parent La$_{2}$NiO$_{4}$ phase, an effect attributed to large octahedral rotations that accommodate intercalated anions in the absence of spacer layer expansion \cite{Wissel2018_La2NiO3F2}. In contrast, we observe both pronounced $c$-axis expansion and suppression of oxygen octahedral rotations following oxidation of Ruddlesden-Popper nickelates (Fig.\ \ref{n2_n3_halforder}). This observation suggests a strong coupling between octahedral symmetry and topotactic expansion in anion-intercalated Ruddlesden-Popper phases. 

Further studies on bulk and powder samples could clarify why oxygen intercalation has not been previously observed in bulk or powder form and elucidate the role of epitaxial strain in stabilizing oxygen-intercalated layered perovskites. Additionally, intercalation of layered oxides has been used as a route to enable exfoliation \cite{Mallouk2000_RP_exfoliation}, suggesting that oxygen-intercalated nickelates could provide a platform to isolate two-dimensional Ruddlesden-Popper nickelate flakes. More broadly, our findings establish ozone annealing as a powerful synthetic tool to induce topotactic oxygen intercalation in layered oxides.

\subsection{Electronic structure}

We examine the electronic structure of oxygen-intercalated nickelates via X-ray absorption spectroscopy (XAS). The O-$K$ edge pre-peak is a sensitive probe of unoccupied 3$d$ states, nickel valence, and degree of nickel-oxygen hybridization \cite{Groot1989}. Figure \ref{fig4_xas_transport}(a) shows the evolution of the O-$K$ edge pre-peak for La$_{n+1}$Ni$_{n}$O$_{3n+1}$ films ($n=1-3$) before and after oxidation. The pristine La$_{2}$NiO$_{4}$ ($n=1$) film exhibits a weak pre-peak feature consistent with a 2+ nickel valence \cite{Kuiper1991PRB}. Upon oxidation, the pre-peak intensity increases and shifts to lower energy — beyond the nickel 3+ reference — indicating hole doping into hybridized nickel 3$d$ - oxygen 2$p$ states. This behavior is supported by nickel-$L_{2}$ edge spectra, which shift to higher energy upon oxidation (Fig.\ \ref{NiL2_xas_xld}), also surpassing the nickel 3+ reference \cite{Mizokawa2000}. Based on these spectral changes and our COBRA-derived oxygen stoichiometry of $\delta \sim 0.7$ in La$_{2}$NiO$_{4+\delta}$, we estimate a nickel valence of approximately 3.4+, consistent with that of a hole-doped Nd$_{2-x}$Sr$_{x}$NiO$_{4}$ ($x=1.4$) reference (Fig.\ \ref{NiL2_xas_srdoped_ref}). For higher-order ($n>1$) compounds, the oxidation-induced shift in O-$K$ and Ni-$L_{2}$ edge spectra decreases with increasing $n$, indicating a reduced degree of oxidation with increasing $n$.

These experimental trends are supported by DFT calculations, which show that the charge transfer energy, $\Delta_{CT} = \epsilon_{d}-\epsilon_{p}$, decreases from approximately $3.0-3.5$ eV in the parent Ruddlesden-Popper nickelates to $1.7-2.0$ eV in the oxygen intercalated compounds. This reduction reflects a substantial enhancement in nickel-oxygen hybridization upon oxidation (Fig.\ \ref{fig:dft}). Large copper-oxygen hybridization is a feature associated with Zhang-Rice singlet physics in high-T$_{c}$ cuprates, positioning oxygen-intercalated nickelates as a potential platform for high-temperature superconductivity. Additionally, DFT shows that oxidation quenches the tetragonal distortion in the nickel-oxygen octahedra (Supplemental Note 2, Fig.\ \ref{fig5_rp_vs_aurivillius}), corroborated by the suppression of the X-ray linear dichroic signal at the nickel-$L_{2}$ edge following oxidation (Supplementary Note 3, Fig.\ \ref{NiL2_xas_xld}). 

\subsection{Electrical transport properties}

Next, we investigate the electrical transport properties of the oxygen-intercalated nickelate films. We begin with La$_{2}$NiO$_{4}$, a single-layer compound isostructural to the high-temperature superconductor La$_{2-x}$Sr$_x$CuO$_4$. In hole-doped La$_{2-x}$Sr$_{x}$NiO$_{4}$, a rich phase diagram emerges, featuring an antiferromagnetic insulating parent phase, charge ordering and charge/spin stripes, but no superconductivity to date. Notably, metallicity only emerges at high doping levels ($x\sim1$) \cite{Shinomori2002JourPhysSocJapan}. As shown in Fig.\ \ref{fig4_xas_transport}b, the parent La$_{2}$NiO$_{4}$ film is semiconducting, consistent with the undoped phase. Following oxidation, the film becomes metallic, exhibiting a residual resistance ratio (RRR) of 4.2. To our knowledge, this is the first observation of metallicity in a single-layer nickelate in the absence of $A$-site cation substitution. The emergence of metallicity in La$_{2}$NiO$_{4+\delta}$ suggests that oxygen intercalation at $\delta\sim0.7$ exceeds the $x\sim1$ hole doping necessary for metallicity in La$_{2-x}$Sr$_x$NiO$_4$. Additionally, oxygen content can be tuned to a `half-oxidized' phase which exhibits two-layer staging, with intercalation of every two rock salt layers, as well as conductivity intermediate between the pristine and `fully-oxidized' phases (Fig.\ \ref{staging}). 

We next turn to the bi-layer compound, La$_{3}$Ni$_{2}$O$_{7}$, recently reported to exhibit superconductivity at 80 K under hydrostatic pressure in bulk form \cite{Sun2023_SCLa327} and, more recently, at ambient pressure in ozone-annealed thin films \cite{Ko2024Nature}. The transport properties are exquisitely sensitive to oxygen content:  metallic for $\delta \sim0$ and insulating for $\delta>0.08$ \cite{Zhang1994_La327}. As shown in Fig.\ \ref{fig4_xas_transport}c, the pristine La$_{3}$Ni$_{2}$O$_{7}$ film exhibits semiconducting behavior, consistent with the oxygen-deficient phase. After oxidation, the film becomes metallic.

The three-layer compound, La$_{4}$Ni$_{3}$O$_{10}$, is known to be metallic, exhibiting a kink in the resistivity near 140 K attributed to charge/spin density wave ordering \cite{Greenblatt1997}. This feature is visible in our pristine La$_{4}$Ni$_{3}$O$_{10}$ film, as shown in Fig.\ \ref{fig4_xas_transport}(d). While the film remains metallic upon oxidation, the resistivity kink disappears, indicating suppression of the density wave transition. This suppression is further corroborated by temperature-dependent Hall coefficient measurements, which also show the disappearance of the associated anomaly (Fig.\ \ref{n3_hall}). In parallel, half-order diffraction peaks, sensitive to octahedral rotations, also diminish with oxidation (Fig.\ \ref{n2_n3_halforder}, suggesting that oxygen intercalation suppresses both density wave order and octahedral rotations. This behavior mirrors the effects of hydrostatic pressure in bilayer and trilayer nickelates, where high-temperature superconductivity emerges following the suppression of density wave order and octahedral tilts \cite{Sun2023_SCLa327,Zhu2024La4310SC}. 

Collectively, these transport results demonstrate that oxygen intercalation dramatically alters the electronic properties. In contrast to fluorinated nickelates \cite{Wissel2018_La2NiO3F2}, oxygen-intercalated nickelates exhibit metallicity, providing a platform for tuning transport properties through spacer layer chemistry, akin to the charge reservoir layer in cuprates. Furthermore, oxygen intercalation provides a route to achieve high levels of hole doping in the absence of chemical substitution. At the same time, oxygen intercalation $\delta$ and cation substitution $x$ provide complementary approaches to hole doping, as demonstrated in the similar superconducting phase diagrams of La$_{2}$CuO$_{4+\delta}$ and La$_{2-x}$Sr$_{x}$CuO$_{4}$ \cite{JorgensenPRB1988, WellsScience1997_La2CuO4+y}. While we have demonstrated the ability to control $\delta$ between half- and fully-oxidized states (Fig.\ \ref{staging}), further work is needed to determine how finely $\delta$ can be controlled in oxygen-intercalated nickelates.

\section{Conclusions}

Here we have uncovered a new family of oxides — oxygen-intercalated layered perovskites — via topotactic oxidation of Ruddlesden–Popper nickelates.  These compounds incorporate approximately 0.7 interstitial oxygen atoms per formula unit, dramatically more than has been observed previously, resulting in a structural intermediate between Ruddlesden–Popper and Aurivillius phases. In addition to a large expansion of the out-of-plane lattice constant, oxygen intercalation suppresses octahedral rotations, a structural feature associated with high-temperature superconductivity in Ruddlesden-Popper nickelates \cite{Sun2023_SCLa327, Zhu2024La4310SC}. The electronic structure exhibits enhanced nickel-oxygen hybridization and pronounced oxygen character at the Fermi energy, characteristic of Zhang-Rice singlet physics in hole-doped cuprates. Furthermore, these phases are metallic, providing a platform for electronic doping through spacer layer chemistry, analogous to charge reservoir layers in cuprates. More broadly, this work provides a synthetic approach for accessing highly oxidized metastable phases across a wide class of layered oxides.

\newpage
\clearpage


%


\vspace{5mm}

\noindent{\Large\textbf{Methods}}\\

\noindent \textbf{Thin film synthesis.} We use ozone-assisted MBE to synthesize the Ruddlesden-Popper nickelate thin films on NdGaO$_{3}$ (110). The MBE synthesis conditions and calibration scheme are described in Refs.\ \cite{PanPRM2022_RPgrowth, Li_YNie2020APL}.

\vspace{5mm}

\noindent \textbf{Topotactic oxidation.} Ozone annealing was carried out using a custom-built furnace equipped with an ozone generator (AX8400, Astex) operating at ambient pressure. The input gas consisted of either 99.999\% pure oxygen or a 99.999\% O$_2$ + 100 ppm N$_2$ mixture. The ozone generator was operated at power levels between 2\% and 20\%, with a constant gas flow rate of 0.7 sccm and an ozone generator pressure of 50 psi provided by a back-pressure regulator. The ozone nozzle was positioned 1.5 cm from the sample surface. Heating and cooling rates were precisely controlled at 20 K/min using a PID controller (Heeg Vacuum Engineering). Typical annealing conditions involved heating the sample to $300 \degree$C for 3 hours. 

\vspace{5mm}

\noindent \textbf{Structural characterization.} Lab-based X-ray diffraction measurements were performed on a Malvern Panalytical Empyrean diffractometer using Cu K$\alpha_{1}$ ($\lambda = 1.5406\ \angstrom$) radiation. Reciprocal space maps (RSMs) were acquired using a PIXcel3D area detector. Lattice constants were determined using Nelson-Riley fitting of the superlattice peak positions. Scanning transmission electron microscopy (STEM) was performed on Thermo Fisher Scientific (TFS) Spectra operated at 200 keV with a convergence semi-angle of 18.9 mrad. Integrated differential phase contrast (iDPC-) STEM was performed using a segmented quadrant dark-field detector. Electron-transparent lamellae were prepared with a TFS Helios 660 focused ion beam (FIB). The oxygen-intercalated nickelates proved highly electron-beam-sensitive: during TEM measurements, they quickly relaxed to the pristine Ruddlesden-Popper phase, as evidenced by a continual decrease in the out-of-plane ($c$-axis) lattice parameter.

\vspace{5mm}

\noindent \textbf{Surface X-ray diffraction and COBRA method} Synchrotron surface X-ray diffraction measurements were conducted at the beamline sector 7-ID-C of the Advanced Photon Source, Argonne National Laboratory, using a six-circle Huber diffractometer configured in Psi-C geometry. The incident X-ray beam, fine tuned to an energy of 17.5 keV ($\lambda = 0.70846\ \angstrom$) using a Si (111) double-crystal monochromator ($\Delta$E/E $\approx$ 1 × 10$^{-4}$), was focused to a 30 $\mu$m (vertical) × 50 $\mu$m (horizontal) beam spot using Kirkpatrick–Baez mirrors, delivering a total photon flux of $\sim$ 3 × 10$^{12}$ photons/s. Bragg rod measurements, including both specular (00L) and off-specular (HKL) reflections, were acquired using an Eiger2 X 500K area detector, with diffraction data collected up to L = 3.9 reciprocal lattice units (r.l.u.). For specular CTR measurements, incident and diffracted angles were kept the same, while off-specular CTR measurements employed a fixed incident angle of 5$^{\circ}$ to maintain the same X-ray footprint. Raw 2D detector images were corrected for background scattering, geometric factors, and non-uniform pixel response. The resulting data were analyzed using the COBRA method, a phase retrieval algorithm optimized for systems with two-dimensional in-plane periodicity and out-of-plane structural variation. Three dimensional electron density maps with sub-$\angstrom$ resolution were reconstructed via an iterative difference map algorithm. Structural models were initialized using the known bulk crystal structures of NdGaO$_{3}$ and RP nickelate phases, as appropriate, and refined to incorporate coherently strained thin-film configurations.

\vspace{5mm}

\noindent \textbf{Transport measurements.} Electrical transport data were taken using the electrical transport option (ETO) on a Quantum Design Physical Property Measurements System (PPMS) equipped with a 14 Tesla magnet. Hall bars were defined by hand using a diamond scribe. Electrical contact was achieved by wire-bonding with aluminum wire to 20 nm palladium contacts deposited via e-beam evaporation. Resistance measurements were taken using an AC 1 $\micro$A current at 18.3 Hz.

\vspace{5mm}

\noindent \textbf{X-ray absorption spectroscopy.} X-ray absorption spectroscopy (XAS) measurements were conducted at the Advanced Light Source, Lawrence Berkeley National Laboratory, on Beamline 6.3.1. The spectra were acquired in total electron yield mode at room temperature. Linearly polarized incident photons were fixed, while the sample was oriented either normal ($I_{x}$) to the beam or at a 30\degree{} grazing incidence angle ($I_{z}$). A geometric correction factor was applied to the grazing incidence signal. The XAS spectra were normalized to the incident X-ray flux, scaled to an intensity of one below the absorption edge, and further normalized across the entire edge. Each spectrum represents an average of 4–8 pairs of measurements taken at normal and grazing incidence angles. The La-$M_{4,5}$ edges were used for energy calibration of the nickel and oxygen spectra. X-ray linear dichroism (XLD) signals were calculated as $I_z - I_x$ ($E \parallel c - E \parallel a,b$) and normalized by $2/3 I_z + I_x$.

\vspace{5mm}

\noindent \textbf{Data Availability}\\
The data that support the findings of this study are available from the corresponding authors upon reasonable request.

\vspace{5mm}

\noindent \textbf{Acknowledgements}\\
We thank Michael Hayward for fruitful discussions. This project was primarily supported by the U.S. Department of Energy, Office of Basic Energy Sciences, Division of Materials Sciences and Engineering, under Award No. DE-SC0021925. This work made use of the Advanced Light Source, a U.S. DOE Office of Science User Facility under contract No. DE-AC02-05CH11231. This research was also performed on APS beam time proposal No.\ 1009892 from the Advanced Photon Source, a U.S. Department of Energy (DOE) Office of Science user facility operated for the DOE Office of Science by Argonne National Laboratory under Contract No. DE-AC02-06CH11357. This work was carried out in part through the use of MIT.nano's facilities and the Harvard University Center for Nanoscale Systems (CNS), a member of the National Nanotechnology Coordinated Infrastructure Network (NNCI), supported by the National Science Foundation under NSF award No.\ ECCS-2025158. Certain commercial equipment, instruments, software, or materials are identified in this paper in order to specify the experimental procedure adequately. Such identifications are not intended to imply recommendation or endorsement by NIST, nor is it intended to imply that the materials or equipment identified are necessarily the best available for the purpose. D.F.S. and, G.A.P., M.B.S., and A.Y.J. acknowledge support from the NSF Graduate Research Fellowship No. DE-SC0021925. and the U.S. Department of Energy, Office of Basic Energy Sciences, Division of Materials Sciences and Engineering, under award No. DE-SC0021925. N.K.T. and A.Y.J. acknowledges support from the Ford Foundation Predoctoral Fellowship. N.K.T. also acknowledges support from the ALS Doctoral Fellowship in Residence. G.A.P. and A.Y.J. acknowledge support from the Paul and Daisy Soros Fellowship for New Americans. A.B.T. acknowledges support from NSF DMR-2323970. S.H.S. and I.E. were supported by the Rowland Institute at Harvard. J.A.M.Mundy acknowledges support from the Packard Foundation and the Gordon and Betty Moore Foundation’s EPiQS Initiative, grant GBMF6760. A.S.B. and H.L. acknowledge NSF grant No. DMR-2323971 and the ASU research computing center for HPC resources. A.B.T. acknowledges support from NSF-DMREF grant No\ DMR-2323970.

\vspace{5mm}

\noindent \textbf{Author contributions}\\
\noindent D.F.S, A.B.T., G.A.P., A.Y.J. and M.B.S. synthesized the thin films via MBE with assistance from C.M.B. and J.A.M. J.K., T.S., C.K.K., D.F.S., A.T., and D.G.S. ozone-annealed the films. H.L. and A.S.B. performed density functional theory calculations. N.K.T., D.F.S., K.-Y.B., and A.T.N. performed X-ray absorption spectroscopy measurements; N.K.T. and A.T.N. analyzed the XAS data. S.H.S. and I.E. performed electron microscopy measurements. D.F.S., K.-Y.B., D.A.W., and H.Z. took synchrotron X-ray diffraction data and performed COBRA analysis. A.J.G. and S.M. performed SIMS measurements. J.A.M. conceived and guided the study. D.F.S. and J.A.M. wrote the manuscript with discussion and contributions from all authors.

\vspace{5mm}

\noindent \textbf{Competing interests}
The authors declare no competing interests.

\vspace{5mm}

\section*{Additional information}

\noindent \textbf{Supplementary information is available} for this paper.

\vspace{5mm}

\noindent \textbf{Correspondence and requests for materials} should be addressed to Julia A. Mundy.

\vspace{5mm}

\noindent \textbf{Reprints and permissions information} is available.

\clearpage
\onecolumngrid

\renewcommand{\thefigure}{S\arabic{figure}}
\setcounter{figure}{0}
\renewcommand{\thetable}{S\arabic{table}}
\setcounter{table}{0}

\makeatletter
\renewcommand{\fnum@figure}{\large \figurename~\thefigure}
\renewcommand{\fnum@table}{\large \tablename~\thetable}
\makeatother

{\large 

\begin{center}
  {\LARGE \textbf{Supplementary Information}} \\[0.5em]
  \large
\end{center}

\vspace{1mm}

\begin{center}
  {\textbf{Supplementary Note 1: Structural characterization}} \\[0.5em]
  \large
\end{center}

\begin{figure}[H]
    \centering
    \includegraphics[width = 0.8\columnwidth]{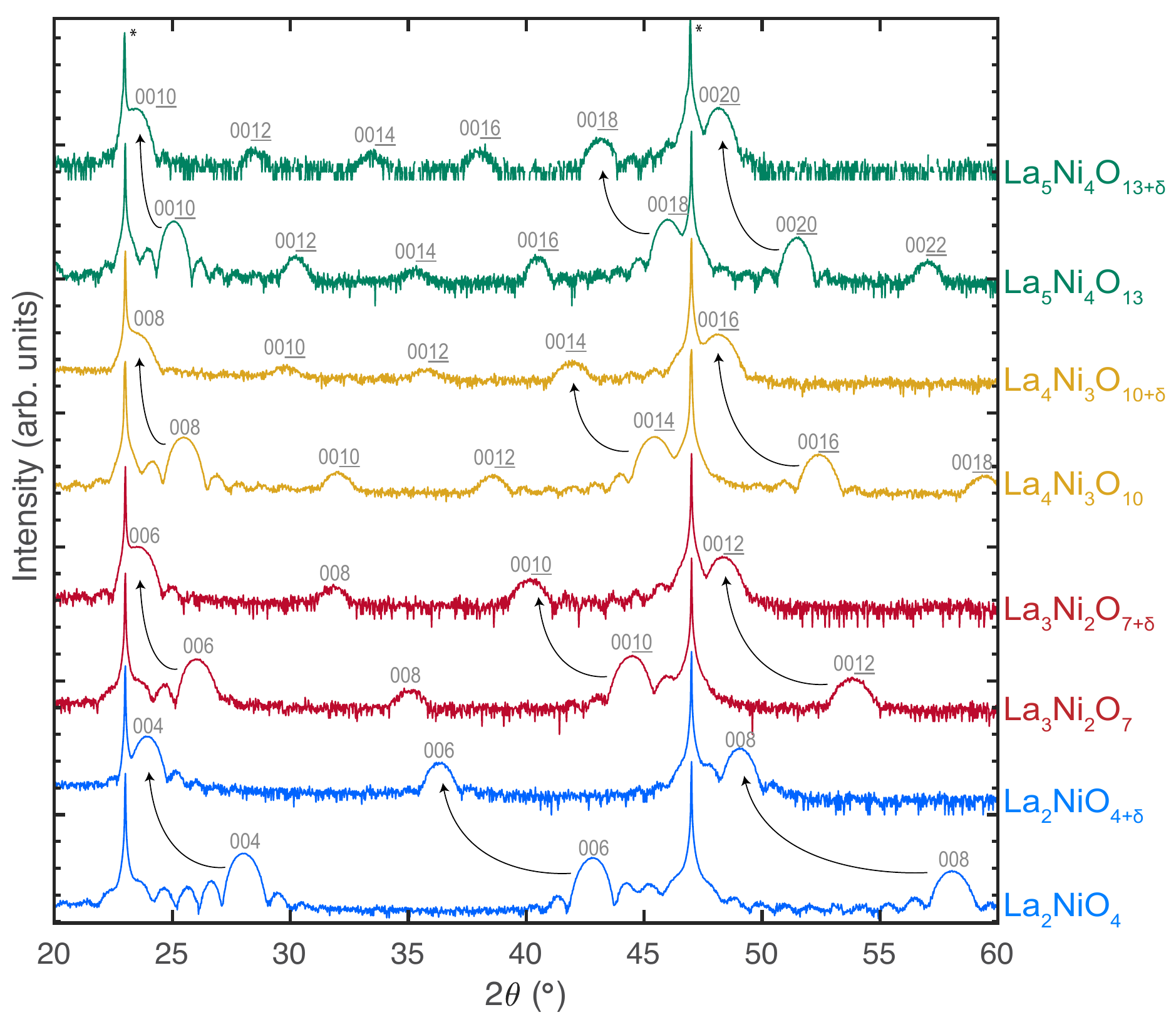}
    \caption{\large \textbf{Lab-based x-ray diffraction scans of La$_{n+1}$Ni$_{n}$O$_{3n+1}$ films before and after ozone treatment.} NdGaO$_{3}$ (110) substrate Bragg peaks are labeled with asterisks. The scans are vertically offset for clarity. Corresponding synchrotron X-ray diffraction data is shown in main text Fig.\ \ref{fig2_00L_scans}.
    }
    \label{lab_xray}
\end{figure}

\begin{table}[H]
\centering
\setlength{\tabcolsep}{12pt} 
\large{
\begin{tabular}{cccccc}
\hline
\makecell{\textbf{compound}} & 
\makecell{\textbf{$c$-axis} \\ \textbf{as-grown (\AA{})}} & 
\makecell{\textbf{$c$-axis} \\ \textbf{oxidized (\AA{})}} & 
\makecell{\textbf{$c$-axis} \\ \textbf{expansion (\AA{})}} & 
\makecell{\textbf{$c$-axis} \\ \textbf{expansion (\%)}} \\
\hline\hline
La$_2$NiO$_{4+\delta}$       & 12.68 $\pm$ 0.11 & 14.93 $\pm$ 0.11 & 2.25 $\pm$ 0.16 & 17.8 $\pm$ 1.3 \\
La$_3$Ni$_2$O$_{7+\delta}$   & 20.63 $\pm$ 0.12 & 22.61 $\pm$ 0.09 & 1.98 $\pm$ 0.15 &  9.61 $\pm$ 0.73  \\
La$_4$Ni$_3$O$_{10+\delta}$  & 28.24 $\pm$ 0.11 & 30.27 $\pm$ 0.10 & 2.03 $\pm$ 0.15 & 7.19 $\pm$ 0.54 \\
La$_5$Ni$_4$O$_{13+\delta}$  & 35.70 $\pm$ 0.14 & 37.86 $\pm$ 0.22 & 2.16 $\pm$ 0.26 & 6.05 $\pm$ 0.72  \\
\hline\hline
\end{tabular}}
\caption{\textbf{Structural parameters for La$_{n+}$Ni$_{n}$O$_{3n+1}$ films before and after oxidation.} CTR and lab-based x-ray diffraction scans for these films are shown in main text Fig.\ \ref{fig2_00L_scans} and Supplemental Fig.\ \ref{lab_xray}, respectively. $c$-axis lattice constants are calculated via Nelson-Riley fits of the synchrotron (0,0,$L$) CTR peak positions \cite{NelsonRiley1945}. Errors are 95\% confidence intervals of the Nelson-Riley fit, as described in Ref.\ \cite{Pan2022PRM}.}
\label{tab:caxis}
\end{table}

\begin{figure}[H]
    \centering
    \includegraphics[width =1\columnwidth]{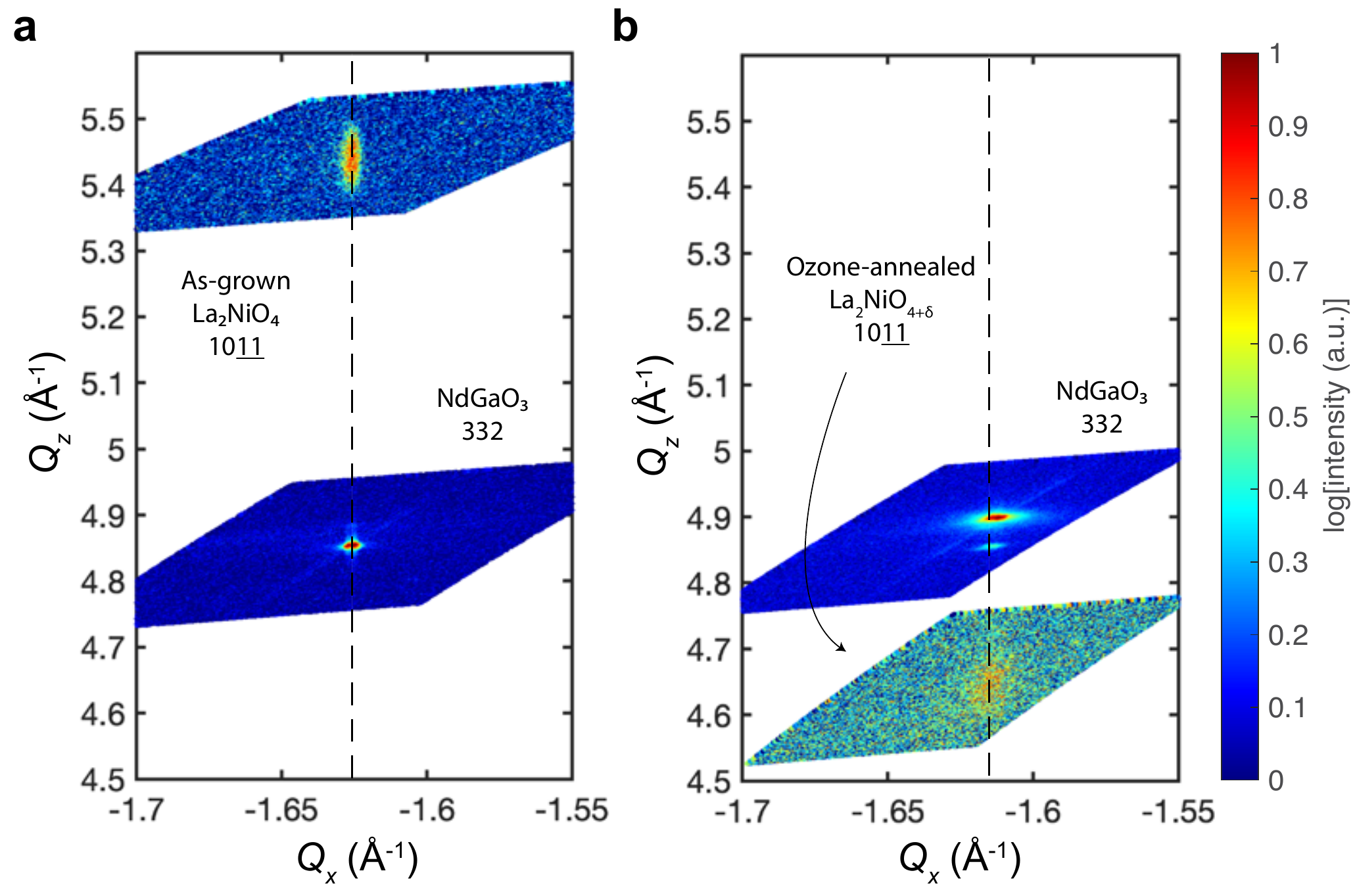}
    \caption{\textbf{Reciprocal space maps of La$_{2}$NiO$_{4}$ before and after oxidation.} \textbf{a, b,} Reciprocal space map of La$_{2}$NiO$_{4}$ / NdGaO$_{3}$ (110) before \textbf{(a)} and after \textbf{(b)} oxidation. The vertical dashed lines highlight that the film and substrate peak positions are aligned in $Q_{x}$, indicating that the film is epitaxially strained to the substrate.
    }
    \label{214_rsm}
\end{figure}

\begin{figure}[H]
    \centering
    \includegraphics[width =1\columnwidth]{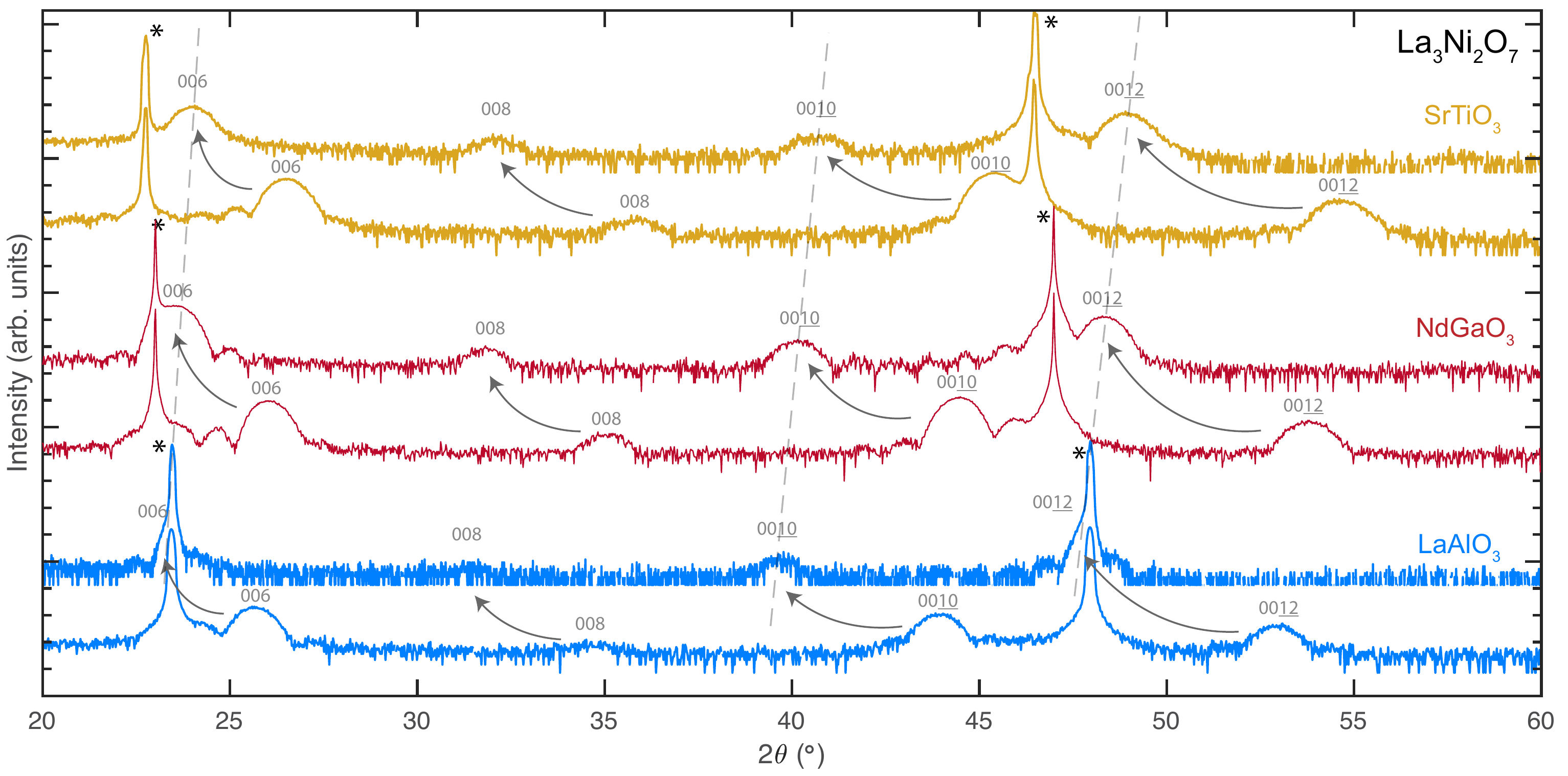}
    \caption{\textbf{Epitaxial strain dependence in as-grown and oxidized La$_{3}$Ni$_{2}$O$_{7+\delta}$ films.} X-ray diffraction scans of as-grown and oxidized La$_{3}$Ni$_{2}$O$_{7+\delta}$ on LaAlO$_{3}$ (100), NdGaO$_{3}$ (110), and SrTiO$_{3}$ (100) substrates. Asterisks indicate substrate Bragg peaks. Dashed lines highlight substrate-dependent Bragg peak positions in the oxidized films, indicating that the films remain epitaxially strained to the substrates following oxidation.
    }
    \label{La327_strain}
\end{figure}

\begin{figure}[H]
    \centering
    \includegraphics[width =1\columnwidth]{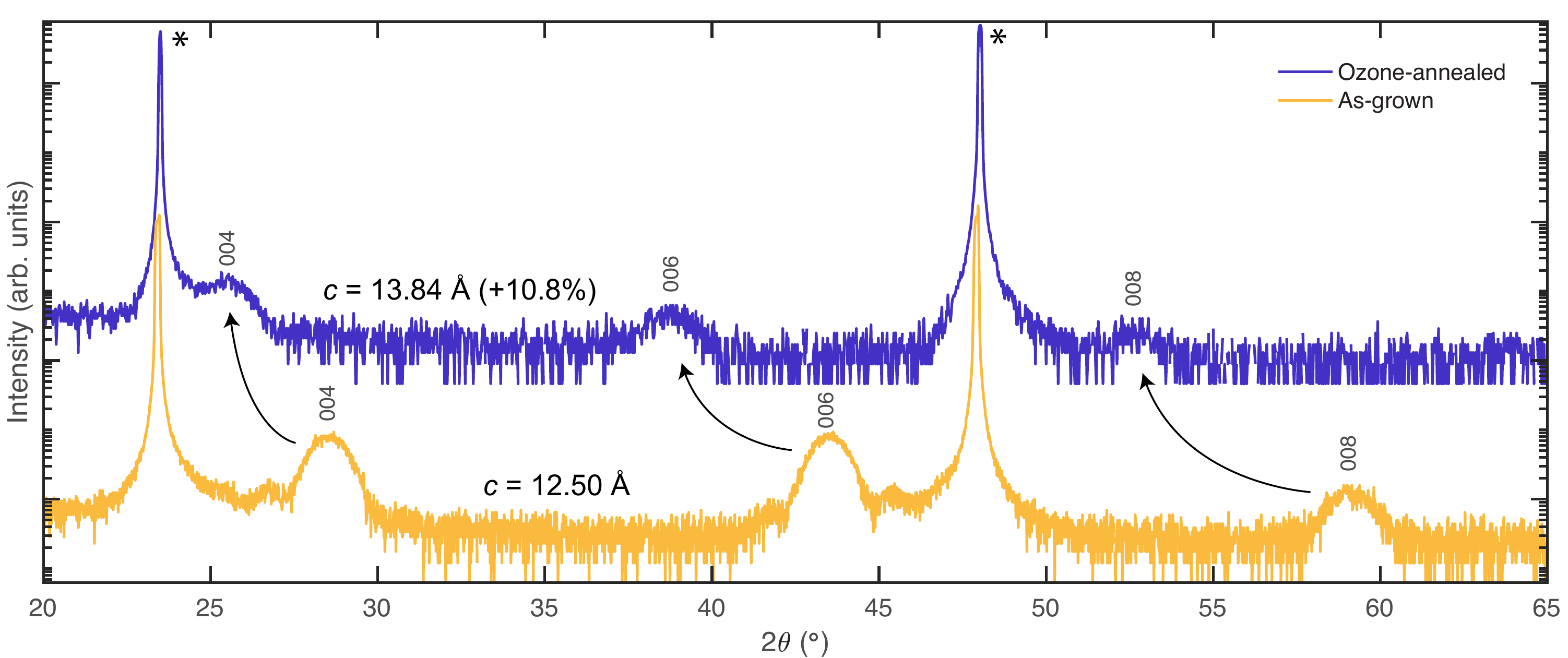}
    \caption{\textbf{Topotactic oxidation of Nd$_{2}$NiO$_{4}$ on LaAlO$_{3}$ (100).} X-ray diffraction scans of as-grown and oxidized Nd$_{2}$NiO$_{4}$ on LaAlO$_{3}$. Asterisks mark the LaAlO$_{3}$ substrate Bragg peaks are marked with asterisks. The $c$-axis lattice constants of the as-grown and oxidized phases are shown, as well as the percentage $c$-axis expansion after oxidation. X-ray absorption spectroscopy data on these samples is shown in Figures\ \ref{NiL2_xas_xld} and \ref{NiL2_xas_srdoped_ref}.
    }
    \label{Nd214_xrd}
\end{figure}

\begin{figure}[H]
    \centering
    \includegraphics[width =1\columnwidth]{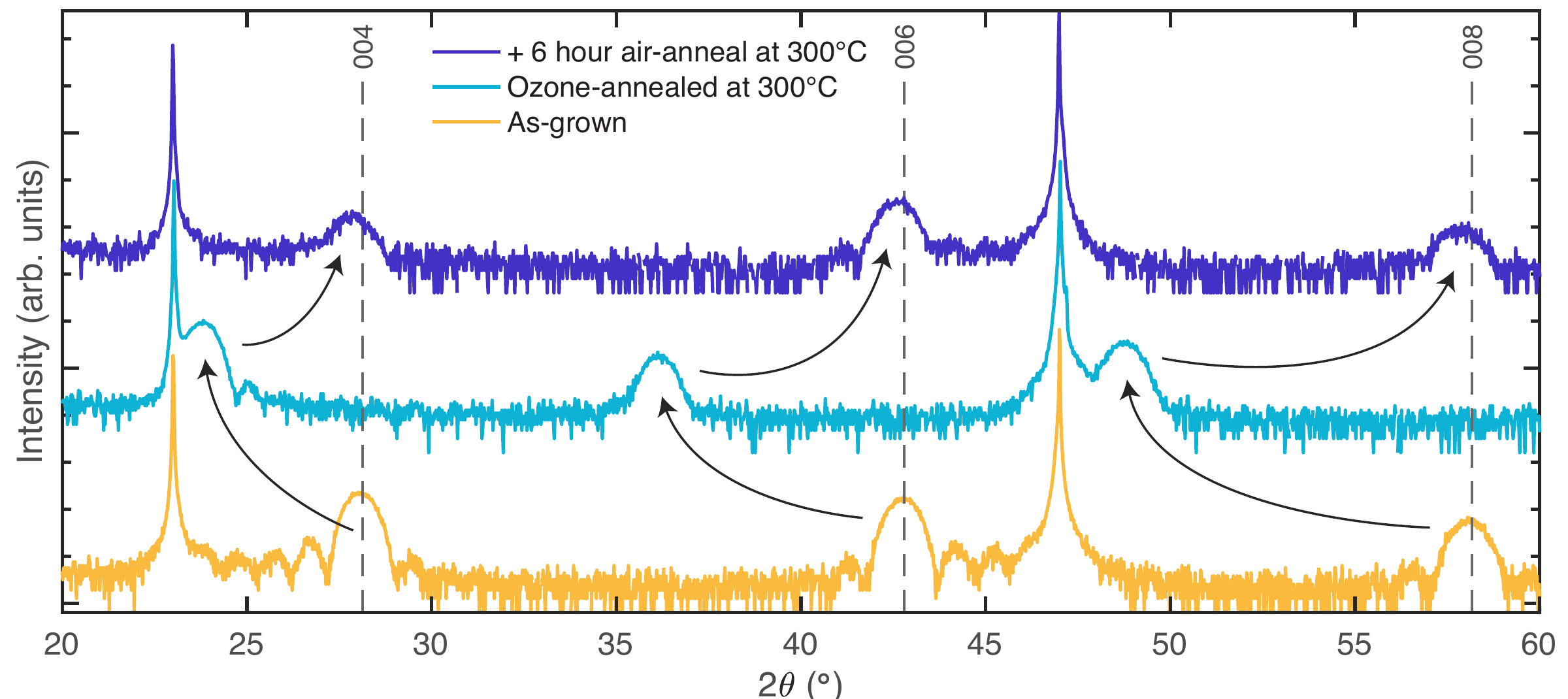}
    \caption{\textbf{Reversibility of topotactic oxidation.} X-ray diffraction scans of a La$_{2}$NiO$_{4}$ / NdGaO$_{3}$ (110) film as-grown, after ozone-annealing, and after subsequent annealing in air. The parent Ruddlesden-Popper structure is recovered after annealing the oxidized film in air, demonstrating the reversibility of the topotactic oxidation reaction. Some structural degradation is evident in the air-annealed film. 
    }
    \label{air_anneal}
\end{figure}

\begin{figure}[H]
    \centering
    \includegraphics[width =1\columnwidth]{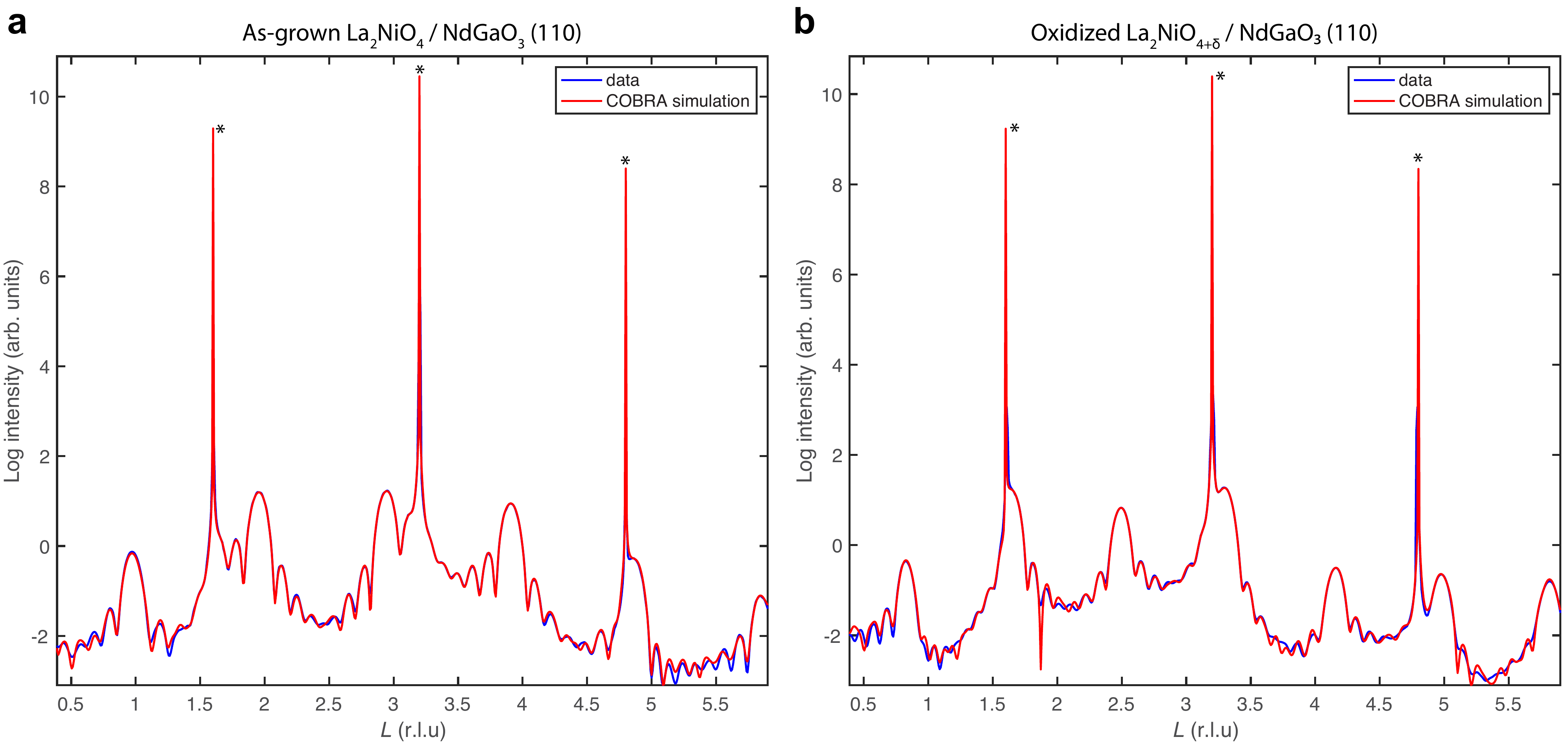}
    \caption{\textbf{COBRA fitting of specular CTR for $n=1$ compound.} \textbf{a, b,} COBRA fits of (0,0,$L$) specular CTR scans of as-grown \textbf{(a)} and oxidized \textbf{(b)} La$_{2}$NiO$_{4}$ / NdGaO$_{3}$ (110). NdGaO$_{3}$ (110) substrate Bragg peaks are labeled with asterisks. The 1-D electron density profiles shown in main text 3a were determined from these fits.
    }
    \label{cobra_00L_fits}
\end{figure}

\begin{figure}[H]
    \centering
    \includegraphics[width =0.8\columnwidth]{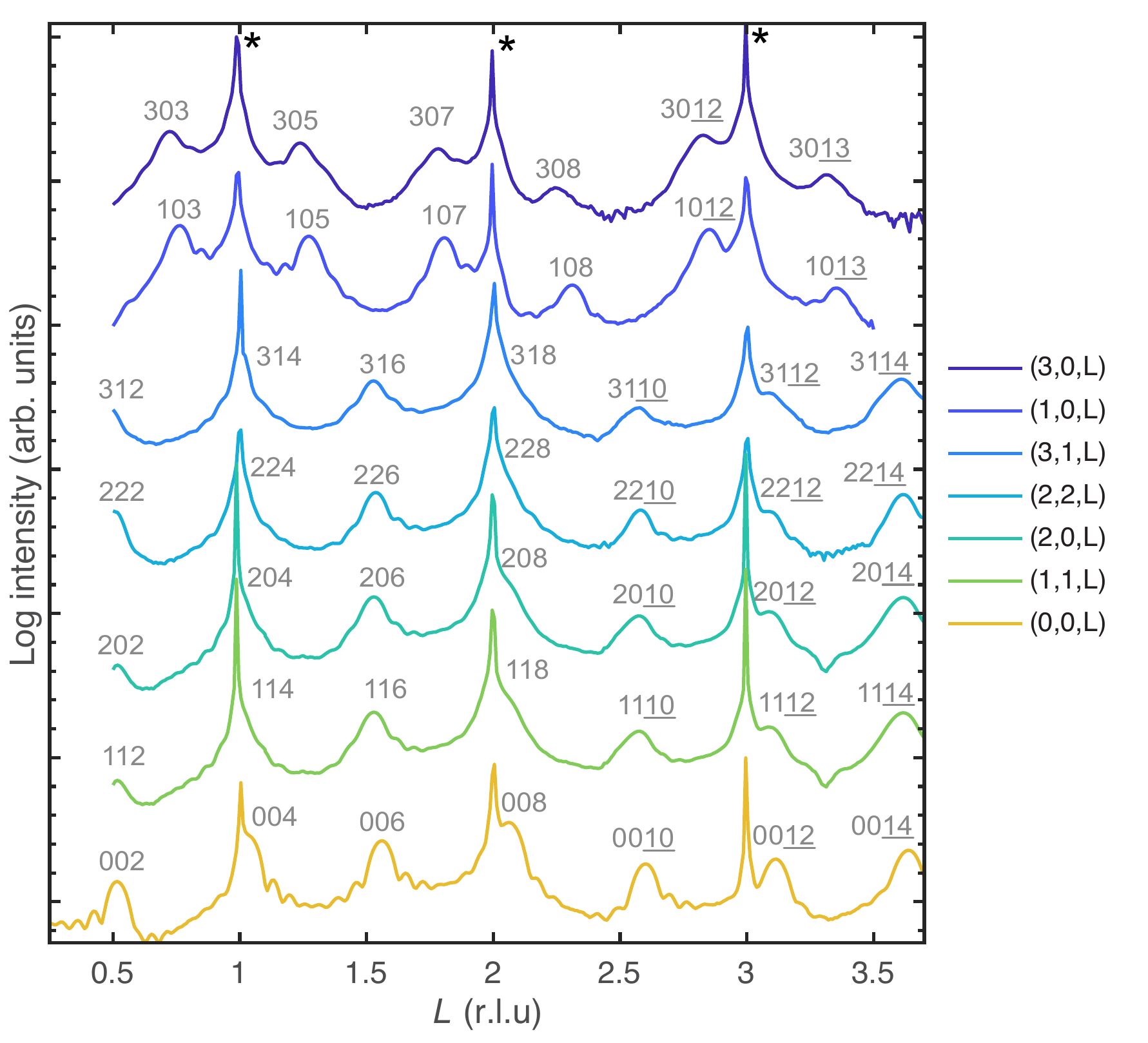}
    \caption{\textbf{Both specular and non-specular CTR scans of oxidized La$_{2}$NiO$_{4+\delta}$}. NdGaO$_{3}$ (110) substrate Bragg peaks are labeled with asterisks. The 3-D electron density maps shown in main text Fig.\ \ref{fig3_cobra}b and c were determined from these CTR scans.
    }
    \label{La215_nonspecular_CTRs}
\end{figure}

\begin{figure}[H]
    \centering
    \includegraphics[width =1\columnwidth]{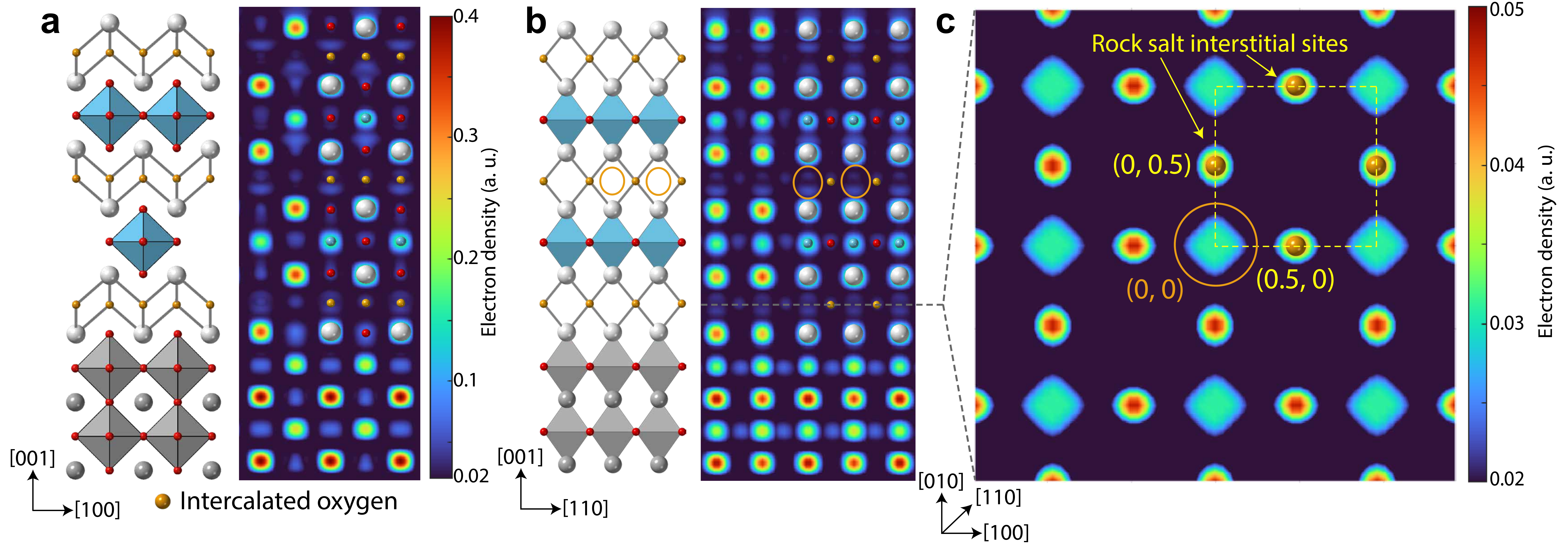}
    \caption{\textbf{3-D COBRA analysis of oxidized La$_{2}$NiO$_{4+\delta}$.} \textbf{a, b,} Projected 2-D electron density maps along the in-plane [010] \textbf{(a)} and [1$\bar{1}$0] \textbf{(b)} directions. The [010] projection is formed by combining cuts at ($a$,$b$) = (0,0) and (0,0.5), while the [1$\bar{1}$0] projected includes cuts at ($a$,$b$) = (0,0), (0,0.5), and (0,1). Gold circles denote electron density localized at ($a$,$b$) = (0,0), as discussed in \textbf{(c)}. \textbf{c,} In-plane 2-D electron density slice within spacer layer, indicated by the dotted line in \textbf{(b)}. Electron density is observed at two distinct sites: (i) rock salt interstitial sites at ($a$,$b$) = (0.5,0) and (0,0.5), and (ii) an unexpected site at ($a$,$b$) = (0,0), which does not correspond to a typical interstitial position in the rock salt structure \cite{JorgensenPRB1989_La2NiO4p18}. We note that the 3-D electron density reflects a statistical average over a macroscopic $\sim$100 $\mu$m$^{2}$ area and contributions from multiple structural domains may appear simultaneously in the electron density maps.
    }
    \label{cobra_110_spacer}
\end{figure}

\begin{figure}[H]
    \centering
    \includegraphics[width =1\columnwidth]{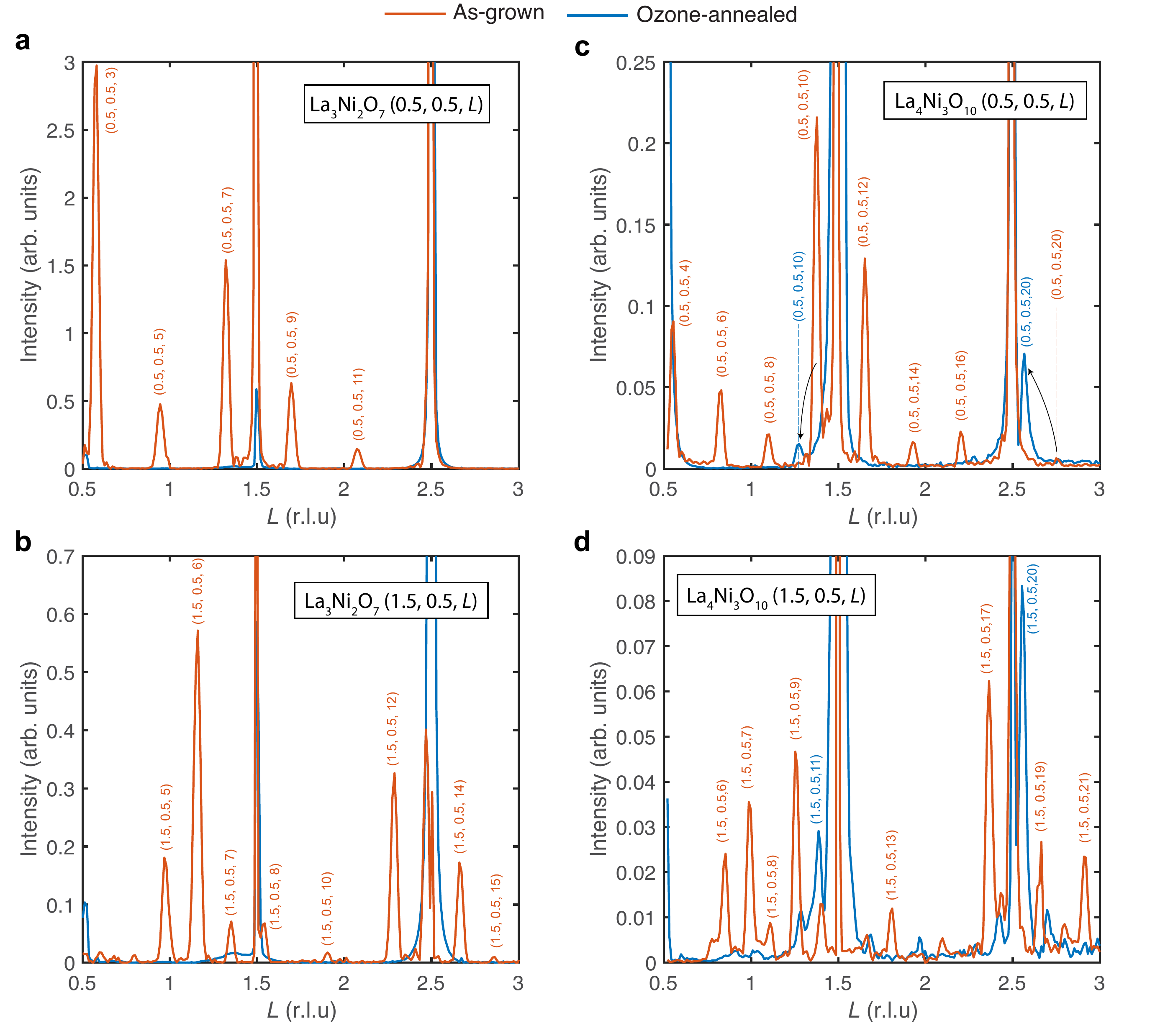}
    \caption{\textbf{Half-order CTR scans probing doubling periodicity oxygen octahedral rotations.} \textbf{a, b,} La$_{3}$Ni$_{2}$O$_{7}$ half-order CTR scans at (0.5,0.5,$L$) \textbf{(a)} and (1.5,0.5,$L$) \textbf{(b)}. \textbf{c, d,} La$_{4}$Ni$_{3}$O$_{10}$ half-order CTR scans at (0.5,0.5,$L$) \textbf{(c)} and (1.5,0.5,$L$) \textbf{(d)}. Peak reflections are labeled. Peaks along the (0.5,0.5,$L$) and (1.5,0.5,$L$) directions are sensitive to octahedral tilting in the in-plane and out-of-plane directions, respectively \cite{May2010PRB,Akamatsu2014_RP_OOR,Yoshida2018_RP_OOR,Bhattacharya2025,Park2025AdvMat}. For La$_{3}$Ni$_{2}$O$_{7}$, the half-order peaks almost entirely disappear after oxidation, indicating quenching of octahedral tilts. In La$_{4}$Ni$_{3}$O$_{10}$, on the other hand, oxidation decreases the half-order peak intensities but not entirely, indicating partial quenching of octahedral rotations. Arrows indicate peak shifts following oxidation. Vertical dotted lines mark certain Bragg peaks that are difficult to label. These results suggest that topotactic oxidation offers a means to modulate oxygen octahedral structural frameworks, a structural degree of freedom linked to high-temperature superconductivity in bi-layer and tri-layer Ruddlesden-Popper nickelates \cite{Sun2023_SCLa327, Zhu2024La4310SC}.
    }
    \label{n2_n3_halforder}
\end{figure}

\clearpage
\begin{center}
  {\textbf{Supplementary Note 2: Electronic structure calculations}} \\[0.5em]
  \large
\end{center}

Electronic band structure calculations were performed using density-functional theory (DFT). We model the intercalated Ruddlesden-Popper (RP) phases, La$_{n+1}$Ni$_{n}$O$_{3n+1+\delta}$, by placing an additional oxygen ($\delta=1$) within the rock salt layer (LaO) which brings the general chemical formula of the series to La$_{n+1}$Ni$_{n}$O$_{3n+2}$. In general, the RP nickelates host a variety of different crystal symmetries, therefore we choose the simplest tetragonal space group symmetry (\textit{P4$_{2}$/mmc}) that can accommodate the additional interstitial oxygen~\cite{lander1989,Rodriguez-Carvajal1991}. To connect to the experiments, we fix the in-plane lattice constants to match the NdGaO$_{3}$ substrate. For the geometry relaxations ($c$-axis and ionic positions), the augmented plane-wave, pseudopotential VASP code~\cite{Kresse:1993bz,Kresse:1996kl, Kresse:1999dk} was used within the generalized gradient approximation of Perdew-Burke-Ernzerhof~\cite{gga_pbe}. A plane-wave energy cutoff of 520 eV and Monkhorst-Pack $k$-grid with dimensions $8\times8\times2$ for $n=1$ and $8\times8\times1$ for $n=2,3$ were used, a Gaussian smearing of 0.1 eV was chosen. With the relaxed crystal geometry, DFT band structure calculations are performed using the all-electron, full-potential WIEN2k code~\cite{Blaha2020wien2k} where the local density approximation is chosen as the exchange-correlation functional. The basis set size was set by $R_{\mathrm{MT}}K_{\mathrm{max}} = 7$ and the muffin-tin radii chosen  (in a.u.) were 2.35, 1.81, and 1.61 for La, Ni, and O, respectively. Brillouin zone integration was performed on a dense $k$-grid of $20\times20\times 5$ for $n=1$ and $20\times20\times3$ for $n=2,3$.

Introducing the interstitial oxygen into the rock salt layer induces significant structural changes. Most notably, the $c$-axis lattice constant undergoes a substantial expansion to accommodate the added charge, as summarized in Table~\ref{tab:dftgeometry}. This expansion qualitatively matches the experimental trend. Achieving quantitative agreement, however, would likely require refining the structural model to account for the same oxygen content and disorder effects. On a more local level, the NiO$_6$ octahedra become quenched, leading to a contraction of the Ni–O bond lengths. The apical Ni–O distances fall in the range of $\sim$1.8–1.9 \AA{} compared to the apical Ni-O distances of $\sim 2.0-2.1$ \AA{} in the parent RPs. The in-plane Ni-O distances remain similar between the two phases
around 1.9 \AA{}. The result is a more symmetric octahedral cage around the nickel atoms in this new family of RPs, in agreement with experiments.

\begin{table}[H]
\centering
\setlength{\tabcolsep}{8pt} 
\large{
\begin{tabular}{ccccc}
\hline
\makecell{\textbf{Parent} \\ \textbf{compound}} & \textbf{$c$-axis} (\AA{}) & \makecell{\textbf{Intercalated}\\  \textbf{compound}} & \textbf{$c$-axis} (\AA{}) & \makecell{\textbf{$c$-axis}\\ \textbf{expansion} (\%)}\\
\hline\hline
La$_2$NiO$_4$   & 12.94 & La$_2$NiO$_{4+1}$ & 14.46 & 11.7 \\
La$_3$Ni$_2$O$_7$ & 20.40 & La$_3$Ni$_2$O$_{7+1}$ & 21.69 & 6.3 \\
La$_4$Ni$_3$O$_{10}$ & 28.05 & La$_4$Ni$_3$O$_{10+1}$ & 29.50 & 5.2 \\
\hline\hline
\end{tabular}}
\caption{\textbf{Lattice constants obtained from DFT for the parent and oxygen-intercalated Ruddlesden-Popper phases.} The in-plane lattice constants have been fixed to match the substrate NdGaO$_{3}$ ($a=3.84$ \AA{}).}
\label{tab:dftgeometry}
\end{table}

\begin{figure}[H]
   \centering
   \includegraphics[width=\columnwidth]{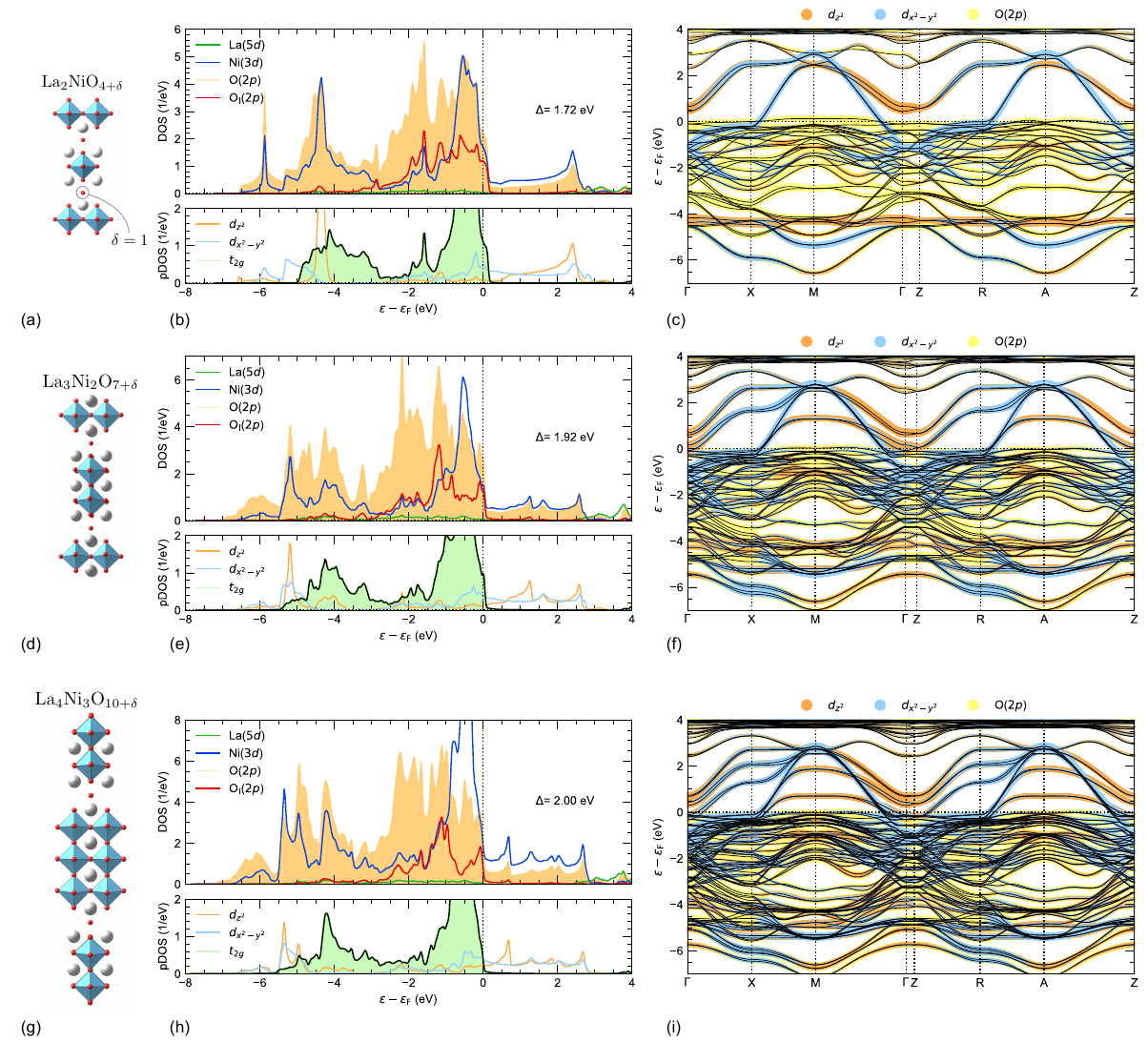}
   \caption{\textbf{DFT electronic structure for intercalated layered nickelate phases for $n=1-3$.} \textbf{a,} Crystal structure for La$_{2}$NiO$_{4+\delta}$ with $\delta=1$ where La, Ni, and O are denoted by gray, blue, and red spheres, respectively. \textbf{b,} Atom- (top) and projected- (bottom) Ni($3d$) density of states (DOS). Note O$_{\mathrm{I}}$ denotes the DOS corresponding to the intercalated oxygen. \textbf{(c,)} Band structure along high-symmetry lines in the Brillouin zone with ``fatband'' character for the Ni-$d_{z^{2}}$ (orange), Ni-$d_{x^{2}-y^{2}}$ (blue), and O($2p$) (yellow) orbital(s). \textbf{(d,e,f,)} Same as \textbf{(a,b,c,)} for La$_{3}$Ni$_{2}$O$_{7+\delta}$ with $\delta=1$. \textbf{(g,h,i,)} Same as \textbf{(a,b,c,)} for La$_{4}$Ni$_{3}$O$_{10+\delta}$ with $\delta=1$.}
   \label{fig:dft}
\end{figure}

Figure~\ref{fig:dft} summarizes the DFT electronic structure for the intercalated RP nickelates with $n=1$–3. Nominally, these compounds should exhibit a highly oxidized nickel state, with formal valences of Ni$^{4+}$ ($d^{6}$), Ni$^{3.5+}$ ($d^{6.5}$), and Ni$^{3.33+}$ ($d^{6.67}$) for $n=1$, 2, and 3, respectively. Thus, in addition to the substantial structural rearrangements, this new family of nickelates is significantly hole-doped with respect to the conventional RP phases. Across all values of $n$, key features of the electronic structure include: (i) strong Ni($3d$) and O($2p$) character at the Fermi level ($\varepsilon_{\mathrm{F}}$), (ii) oxygen-hole character indicative of ligand-hole physics, and (iii) enhanced transition-metal–ligand hybridization, as reflected in the decreased charge-transfer energy $\Delta = \varepsilon_{d} - \varepsilon_{p}$. Estimated from band centroids, $\Delta$ values are 1.72, 1.92, and 2.00 eV for $n=1$, 2, and 3, respectively. The values of the charge-transfer energy are significantly reduced with respect to the parent RP phases where  $\Delta$ $\sim 3-3.5$ eV \cite{labollita2024}. This is directly related to the significant increase in the pre-peak intensity at the O-K x-ray absorption edge shown in Fig.~\ref{fig4_xas_transport}a. The atom-resolved density of states is reminiscent of the itinerant Zhang-Rice singlet (ZRS) physics observed in hole-doped cuprates, characterized by a small charge-transfer energy and significant oxygen character at $\varepsilon_{\mathrm{F}}$. The low-energy nickel states exhibit strong mixing between the $t_{2g} = \{xy, xz, yz\}$ and $e_{g} = \{x^{2}-y^{2}, z^{2}\}$ orbitals. Given the strong tendency of nickel towards a $2+$ oxidation state, it is not surprising to observe extensive multi-orbital character across the full Ni($3d$) manifold, with holes predominantly residing on the surrounding oxygen ligands. Interestingly, these oxygen holes originate not from the intercalated site within the rock salt layer, but primarily from the octahedral oxygen atoms. 

Compared to the parent RP nickelates, the bands with mostly Ni-$e_{g}$ character exhibit similarities and differences. The Ni-$d_{x^{2}-y^{2}}$ bands (light blue color) maintain a similar dispersion as those of the parent compounds, while the Ni-$d_{z^{2}}$ (orange color) undergo a significant transformation. Quantum confinement within the perovskite-like layers results in the emergence of molecular orbitals of Ni-$d_{z^{2}}$ character. These molecular orbitals form bonding and anti-bonding combinations for $n=2$ and bonding, non-bonding, and anti-bonding combinations for $n=3$. For the intercalated compounds, these molecular orbitals remain mostly unoccupied as shown in the Ni($3d$)-projected density of states and band structures. This indicates a substantial alteration to the low-energy physics in these materials with respect to the conventional RPs. Given that the parent compounds become superconducting under pressure, this new family of intercalated RP provides a new materials platform to understand the nature of superconductivity in octahedral coordinated nickelates. 

\clearpage
\begin{center}
  {\textbf{Supplementary Note 3: X-ray absorption spectroscopy}} \\[0.5em]
  \large
\end{center}

The Ni-$L_{2}$ edge probes $2p\rightarrow3d$ transitions and is a powerful tool for determining the nickel valence state and unoccupied $d$-states \cite{Fink1985}. Polarized incident light generates orbital selectivity in the Ni-$L_{2}$ spectra, now sensitive to $E \parallel c$ to $2p\rightarrow3d_{z^{2}}$ and $E \parallel a,b$ ($I_{x}$) to $2p\rightarrow3d_{x^{2}-y^{2}}$ transitions \cite{Pellegrin1995PRB}. First, we assess the polarization-averaged spectra, $\frac{1}{2}(I_z+I_x)$, in Fig.\ \ref{NiL2_xas_xld}a. The as-grown film exhibits a Ni-$L_{2}$ spectrum consistent with La$_{2}$NiO$_{4}$ and a 2+ nickel valence. After oxidation, the Ni-$L_{2}$ peaks shift to higher energy, surpassing the Ni$^{3+}$ reference (Figs. \ref{NiL2_xas_xld} and \ref{NiL2_xas_srdoped_ref}). The energy shift following oxidation decreases with increasing $n$, mirroring trends in the O-$K$ pre-peak (Fig.\ \ref{fig4_xas_transport}a) and $c$-axis expansion (Fig.\ 2c). Furthermore, the Ni-$L_{2}$ edge peak energy is largest in the oxidized $n=1$ compound and decreases with increasing $n$.

A large x-ray linear dichroic (XLD) signal, $I_z - I_x$, is expected when the degeneracy between the $d_{x^{2}-y^{2}}$ and $3d_{z^{2}}$ ($e_{g}$) energy levels is broken. In Nd$_{2}$NiO$_{4}$, a compound known to be in a high-spin 3$d^{8}$ state with one electron in the  $d_{x^{2}-y^{2}}$ and $3d_{z^{2}}$ orbitals, a large XLD signal is observed (Fig.\ \ref{NiL2_xas_xld}b), driven by a tetragonal distortion \cite{Pellegrin1995PRB,kuiper1998}. Following oxidation, the XLD signal significantly decreases. The loss of XLD signal upon oxidation could be driven by the quenching of the degeneracy-breaking tetragonal distortion in Nd$_{2}$NiO$_{4}$: DFT calculations show that the apical Ni-O bond lengths decrease from $\sim$ 2.0-2.1 $\angstrom$ in the parent RP to $\sim$ 1.8-1.9 $\angstrom$ in the oxidized compound (Supplementary Note 2). In the parent compound, apical oxygens are bonded to rare-earth atoms in the spacer layer in the adjacent monolayer (Fig.\ \ref{fig5_rp_vs_aurivillius}a). However, in the oxidized structure, the intercalated species can handle inter-layer bonding, while the apical oxygens can be mostly bonded within the perovskite layer (Fig.\ \ref{fig5_rp_vs_aurivillius}b). The suppressed dichroism could also have an epitaxial strain origin: for example, an XLD signal appears only in $R$NiO$_{3}$ ($d^{7}$) thin film form under epitaxial strain, which breaks the $e_{g}$ degeneracy \cite{Freeland2011,Tung2013_RNiO3_strain,Wu2013}. However, due to the lack of bulk crystal data for the oxygen-intercalated nickelates, we cannot determine the nominal epitaxial strain state.

\newpage

\begin{figure}[H]
   \centering
   \includegraphics[width=0.8\columnwidth]{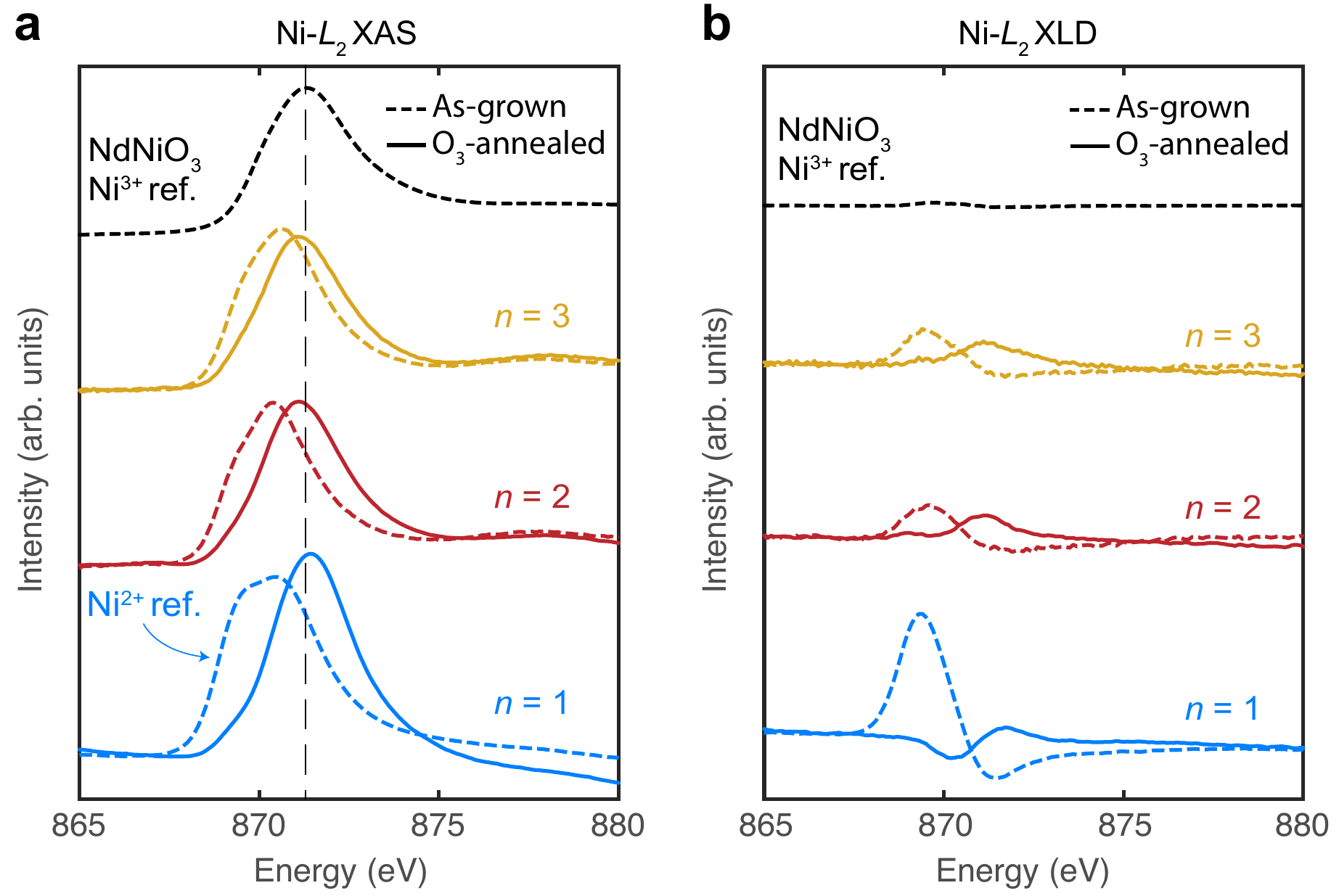}
   \caption{\textbf{Nickel-$L_{2}$ edge spectroscopy of as-grown and oxidized $n=1-3$ films.} A NdNiO$_{3}$ spectrum is shown as a nickel 3+ reference. The spectra are vertically offset for clarity. The $n=1$ film is Nd$_{2}$NiO$_{4}$ / LaAlO$_{3}$ (X-ray diffraction data in Fig.\ \ref{Nd214_xrd}); the $n=2$ and $n=3$ films are lanthanum nickelates on NdGaO$_{3}$ (X-ray diffraction data in Fig.\ \ref{lab_xray}).}
   \label{NiL2_xas_xld}
\end{figure}

\begin{figure}[H]
   \centering
   \includegraphics[width=0.6\columnwidth]{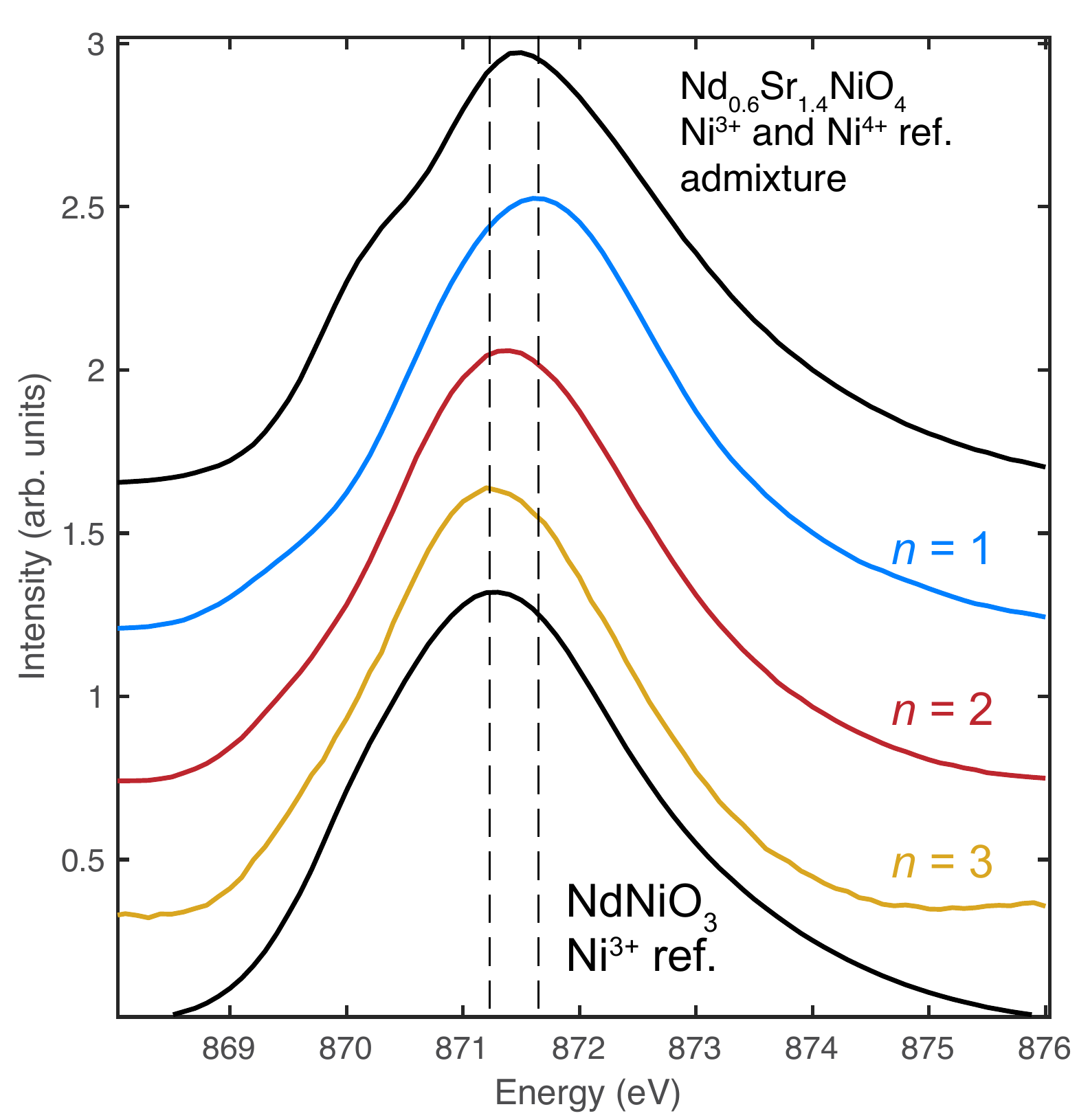}
   \caption{\textbf{Nickel-$L_{2}$ edge spectroscopy of oxidized $n=1-3$ films.} The vertical dotted lines mark the peak positions of the NdNiO$_{3}$ and oxidized $n=1$ spectra. Reference spectra from NdNiO$_{3}$ (Ni$^{3+}$) and Nd$_{0.6}$Sr$_{1.4}$NiO$_{4}$ (nominally Ni$^{3.4+}$) are shown \cite{Taylor2025}. The spectra are vertically offset for clarity. The $n=1$ film is Nd$_{2}$NiO$_{4}$ / LaAlO$_{3}$; the $n=2$ and $n=3$ films are lanthanum nickelates on NdGaO$_{3}$.}
   \label{NiL2_xas_srdoped_ref}
\end{figure}

\clearpage
\begin{center}
  {\textbf{Supplementary Note 4: Transport characterization}} \\[0.5em]
  \large
\end{center}

\begin{figure}[H]
   \centering
   \includegraphics[width=0.7\columnwidth]{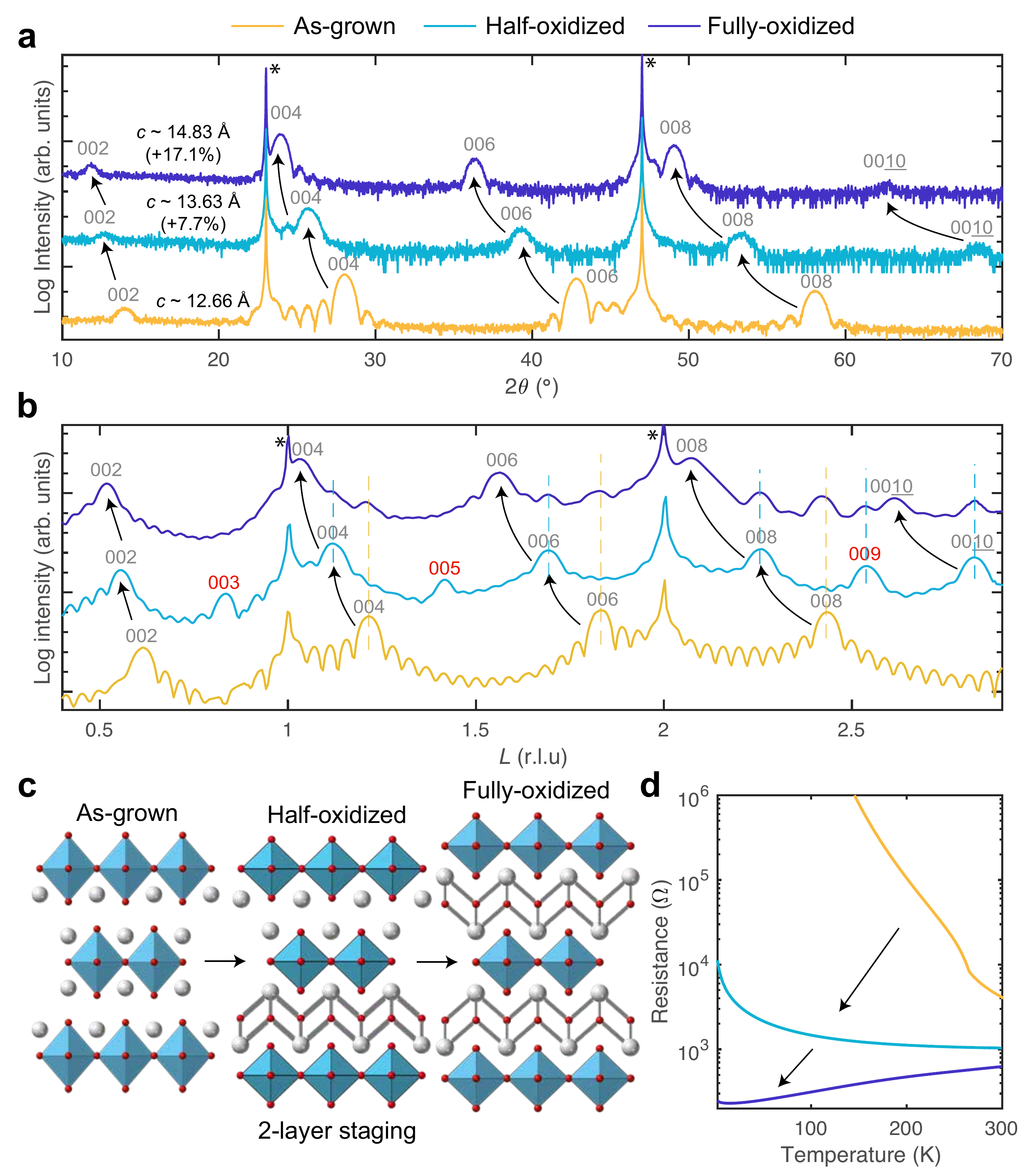}
   \caption{\textbf{Tuning oxygen content in La$_{2}$NiO$_{4+\delta}$.} \textbf{a,} Lab-based X-ray diffraction scans of as-grown, half-oxidized, and fully-oxidized La$_{2}$NiO$_{4}$ films on NdGaO$_{3}$. Asterisks denote NdGaO$_{3}$ (110) Bragg peaks. The $c$-axis lattice constant and percentage increase in $c$-axis relative to the pristine film are noted above each scan. \textbf{b,} Synchrotron X-ray diffraction scans of the same films in \textbf{(a)}. We highlight in red additional odd-order peaks in the half-oxidized film that were not visible by lab X-ray diffraction in \textbf{(a)}. These odd-order peaks likely arise due to `2-layer staging' where every second rock salt layer is intercalated, as illustrated in \textbf{(c)}. Furthermore, the `fully-oxidized' film exhibits peaks from both the half-oxidized and pristine film, marked by colored vertical dotted lines. Note that the odd-order peaks are mostly absent in the `fully-oxidized' film. \textbf{c,} Schematic crystal structures of the as-grown, half-oxidized, and fully-oxidized La$_{2}$NiO$_{4}$ films. \textbf{d,} Temperature-dependent resistance measurements of the three films, showing a progressive increase in conductivity with oxidation, ultimately giving rise to metallicity in the fully-oxidized film.}
   \label{staging}
\end{figure}

\begin{figure}[H]
   \centering
   \includegraphics[width=0.6\columnwidth]{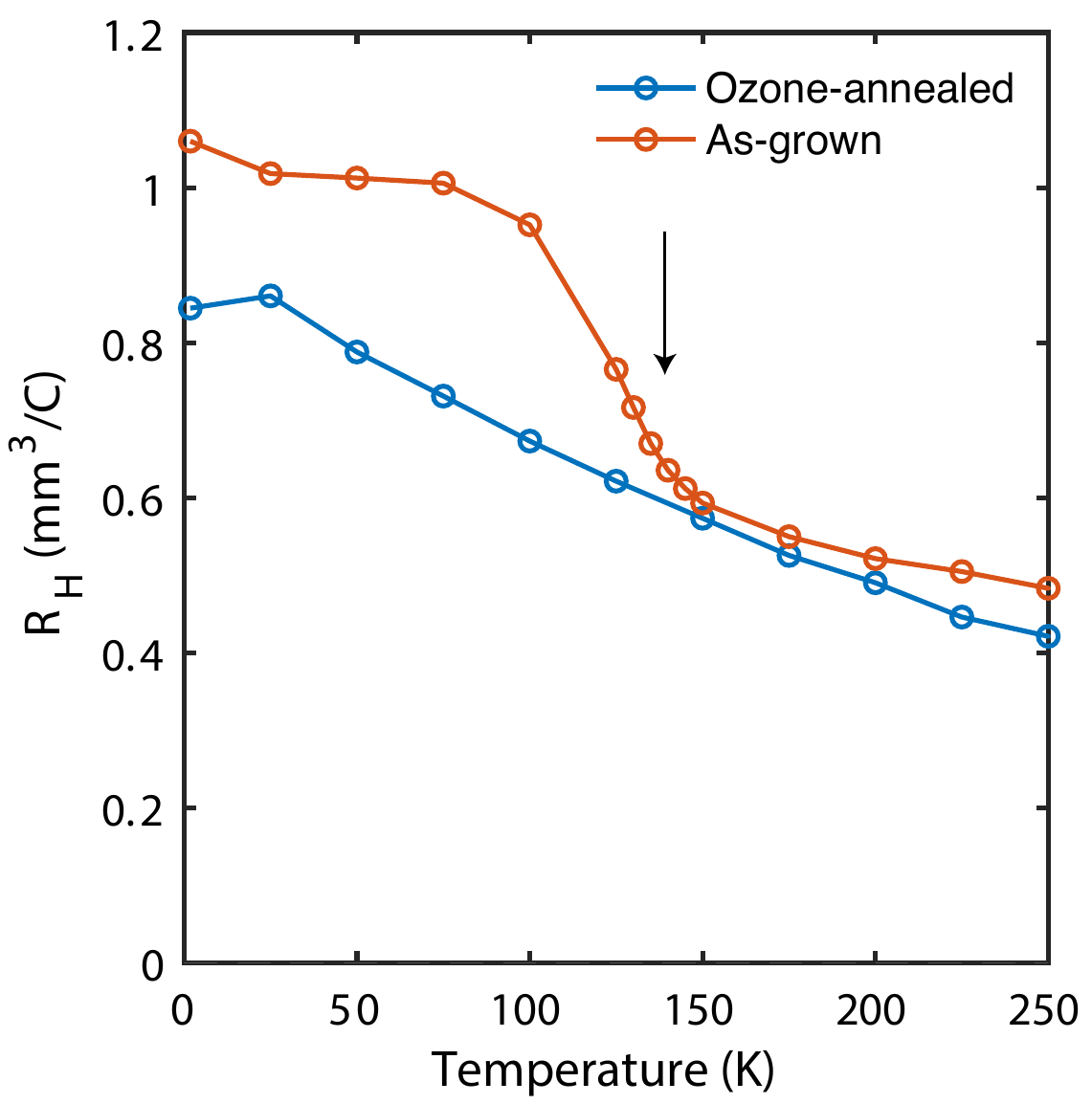}
   \caption{\textbf{Transport characterization of the charge/spin density wave transition in pristine and oxidized La$_{4}$Ni$_{3}$O$_{10}$.} Temperature-dependent Hall coefficients for as-grown and ozone-annealed La$_{4}$Ni$_{3}$O$_{10}$. The arrow marks the charge/spin density wave transition at $\sim140$ K. Temperature-dependent resistivity data is shown in main text Fig.\ \ref{fig4_xas_transport}d.}
   \label{n3_hall}
\end{figure}

\clearpage
\begin{center}
  {\textbf{Supplementary References}} \\[0.5em]
  \large
\end{center}


%


\end{document}